\documentclass[english]{revcoles}
\usepackage[a4paper]{geometry}
\usepackage{graphicx}
\usepackage[textwidth=8em,textsize=small]{todonotes}
\usepackage{amsmath}
\usepackage{amssymb}
\usepackage{amsfonts}
\usepackage{color}
\usepackage{flafter}
\usepackage{url}
\usepackage{enumerate}
\usepackage{multicol}
\usepackage{multirow}
\usepackage[T1]{fontenc}
\usepackage{float}
\usepackage{subfigure}
\usepackage{rotating}
\usepackage{comment}
\usepackage{cancel}
\usepackage{booktabs}

\def\exp#1{{\rm exp}{#1}}
\def\frac#1#2{{{#1}\over{#2}}}

\def\le{\left}
\def\ri{\right}

\def\prop{\propto}
\DeclareMathOperator*{\tr}{tr}

\DeclareMathOperator*{\expit}{expit}

\DeclareMathOperator*{\argmin}{arg\,min}

\newcommand\simiid{\mathrel{\overset{\makebox[0pt]{\mbox{\normalfont\tiny\sffamily iid}}}{\sim}}}
\newcommand\simind{\mathrel{\overset{\makebox[0pt]{\mbox{\normalfont\tiny\sffamily ind}}}{\sim}}}

\newcommand{\pr}[1]{\textsf{Pr}\left[#1\right]}
\newcommand{\expec}[1]{\textsf{E}\left[#1\right]}

\newcommand{\coefvar}[1]{\textsf{CV}\left[#1\right]}
\newcommand{\diag}[1]{\text{diag}\left[#1\right]}

\newcommand{\ex}[1]{\exp{ \left\{ #1 \right\}}}
\newcommand{\quo}[1]{\textquotedblleft#1\textquotedblright}

\def\I{\mathbf{I}}

\def\Q{\mathbf{Q}}

\def\U{\mathbf{U}}\def\uv{\boldsymbol{u}}

\def\X{\mathbf{X}}\def\xv{\boldsymbol{x}}
\def\Y{\mathbf{Y}}

\def\phiv{\boldsymbol{\phi}}
\def\etav{\boldsymbol{\eta}}
\def\tev{\boldsymbol{\theta}}

\def\bev{\boldsymbol{\beta}}
\def\omev{\boldsymbol{\omega}}

\def\be{\beta}
\def\al{\alpha}
\def\sig{\sigma}
\def\si{\sigma}
\def\ome{\omega}
\def\kap{\kappa}
\def\LAM{\mathbf{\Lambda}}
\def\UPS{\mathbf{\Upsilon}}

\def\Dir{\small{\mathsf{Dir}}}

\def\Ber{\small{\mathsf{Ber}}}

\def\Nor{\small{\mathsf{N}}}

\def\Gamd{\small{\mathsf{Gam}}}
\def\IGamd{\small{\mathsf{IGam}}}

\def\trans{\textsf{T}}
\def\reals{\mathbb{R}}
\def\zerov{\boldsymbol{0}}
\def\rest{\textsf{rest}}

\def\DIC{\textsf{DIC}}
\def\WAIC{\textsf{WAIC}}
 
\newcommand{\citet}{\citeasnoun}
\newcommand{\citep}{\cite} 
 
\begin{document}

\title[maintitle = A Review of Latent Space Models for Social Networks,
       secondtitle = Una revisi\'on de Modelos de Espacio Latente para Redes Sociales,
       shorttitle = A review of Latent Space models
]

\begin{authors}
\author[firstname = Juan,
        surname = Sosa,
        numberinstitution = 1,
        affiliation = Professor,
        email = jcsosam@unal.edu.co]
\author[firstname = Lina,
        surname = Buitrago,
        numberinstitution = 1,
        affiliation = Professor,
        email = labuitragor@unal.edu.co]
\end{authors}

\begin{institutions}
     \institute[subdivision = {Departamento de Estad\'stica},
                division = Facultad de Ciencias,
                institution = Universidad Nacional de Colombia,
                city = Bogot\'a,
                country = Colombia]
\end{institutions}

\begin{mainabstract}
In this paper, we provide a review on both fundamentals of social networks and latent space modeling. The former discusses important topics related to network description, including vertex characteristics and network structure; whereas the latter articulates relevant advances in network modeling, including random graph models, generalized random graph models, exponential random graph models, and social space models. We discuss in detail several latent space models provided in literature, providing special attention to distance, class, and eigen models in the context of undirected, binary networks. In addition, we also examine empirically the behavior of these models in terms of prediction and goodness-of-fit using more than twenty popular datasets of the network literature.
\keywords{Bayesian Inference, Markov chain Monte Carlo, Latent Space Model, Social Networks}
\end{mainabstract}

%

\section{Introduction}

The study of information that emerges from the interconnectedness among autonomous elements in a system (and the elements themselves) is extremely important in the understanding of many phenomena. The structure formed by these elements (individuals or actors) and their interactions (ties or connections) is commonly known as a graph, social network, or just \textit{network}. Examples of networks are common in many research areas including: Finance (studying alliances and conflicts among countries as part of the global economy), social science (studying interpersonal social relationships and social schemes of collaboration such as legislative cosponsorship networks), biology (studying arrangements of interacting genes, proteins or organisms), epidemiology (studying the spread of a infectious disease), and computer science (studying the Internet, the World Wide Web, and also communication networks), among many others. Just a few examples are enough to see that both entities and connections in networks are varied and diverse, ranging from people to organizations, and friendship to communication, respectively.

Since the mid 90s there has been an increasing development of statistical methods aiming to improve our understanding of how actors' attributes and relations affect the overall structure and behavior of a system. To that end statistical methods essentially aim to do three things. First, to summarize the patterns that characterize the structure of a network along with its individual entities. Second, to create (stochastic) models that provide a way to explain the process under which a network came to be as it is. And third, to predict missing or future relations taking into account the structural properties of the network and the local rules governing its actors. In contrast to a vast quantity of deterministic methods developed in the physics literature, the implementation of statistical models allow us to report measures of uncertainty associated with parameter estimates and predictions.

This review is structured as follows: Sections \ref{sec_fundamentals} and \ref{sec_properties_of_networks} review fundamental concepts on networks including basic definitions and networks topology. Section \ref{sec_social_space_models} provides details about network modeling, paying special attention to latent space models. Section \ref{sec_computation} presents our approach to Bayesian inference through Markov chain Monte Carlo methods. Section \ref{sec_illustrations} discusses details about distance, class and eigen models, including important properties, prior elicitation, and applications. Finally, Section \ref{sec_discussion} summarizes our main remarks.

\section{Fundamentals}\label{sec_fundamentals}

Generally speaking, network data consists of a set of actors, variables measured on such actors (nodal attributes) and variables measured on pair of actors (dyads). This specific type of data in its simplest form comes in the form of a dichotomous variable indicating the presence or absence of a connection of interest (e.g., friendship, collaboration, alliances and conflicts, and so forth) between a pair of actors: This is known as a binary network. Also, it is quite common to find networks in which edges are equipped with weights (e.g., the amount of time spent together between individuals, costs of transactions between companies, number of conflicts between countries, distance between objects, and so forth) characterizing the corresponding connection between a pair of actors. Such kind of networks are known as weighted or valued networks.

It is also frequent to characterize relations as undirected or directed. An undirected (symmetric) relation has one and only one value per pair of actors; on the other hand, a directed (asymmetric) relation has two values per pair of actors, one value representing the perspective of each pair member. Accordingly, a network is said to be an undirected network if every relation in it is undirected, and it is called directed network or digraph otherwise. Examples of directed networks include the network of citations between academic papers and the network of email messages between coworkers, since each relation is unidirectional. On the other hand, examples of undirected networks include the network of friendship relations and the network of sexual contact between individuals, since there is no directionality implicit in the relation.

A binary network is often represented as a graph in which vertices (nodes) correspond to the actors and edges (ties or links) correspond to the connections between dyads. Another useful way of representing network data is through a matrix commonly known as adjacency matrix or sociomatrix. For binary networks with $I$ nodes, the adjacency matrix $\Y = [y_{i,i'}]$ is an $I \times I$ binary matrix such that $y_{i,i'} = 1$ if there is a link from node $i$ to node $i'$, and $y_{i,i'}=0$ otherwise. Analogously, the adjacency matrix of a weighted network is defined in such a way that $y_{i,i'}$ is equal to the corresponding weight associated with the relation from node $i$ to node $i'$, and it is equal to zero otherwise. The main diagonal of an adjacency matrix is full of structural zeros if edges connecting nodes to themselves are not allowed in the network. Note that the adjacency matrix of an undirected network has to be symmetric; similarly, the adjacency matrix of a directed network is possibly asymmetric.

From a statistical perspective, tools and methods for the analysis of network data can be classified according to three main categories, namely, descriptive methods, modeling and inference methods, and processes methods. First, descriptive methods aim to visualize and numerically characterize the actors and the overall structure of a network. Second, modeling and inference methods aim to explain how a network might have arisen. And third, process methods aim to study how interactions influence actors' attributes. Broadly speaking, in this paper we mostly review methods within the first and second categories.

\section{Network description}\label{sec_properties_of_networks}

Visualization and description are fundamental processes when studying the main features of a network. Graphical techniques and summary quantities, many of them graph-theoretic in nature, have been designed in order to characterize the role of the actors in a network (by describing their relative importance and that of their relations) and the structural patterns of the system (by describing aspects of the network itself such as cohesion, connectivity, assortativity, among many others).

Identifying structural attributes in a network is of great importance because they lead to dependencies in the data. That is why taking into account such dependencies is tremendously important when developing statistical models for network analysis (see Section \ref{sec_statistical_models_for_networks}). Even though the concepts presented below are easily extended to directed networks, for simplicity we devote the discussion principally to undirected, binary networks. Two classic introductory books about network fundamentals and methods are \citet{wasserman-1994} and \citet{scott-2000-social}. More contemporaneous reviews on network properties and measure summaries can be found in \citet{kolaczyk-2009} and \citet{newman-2010}.

In what follows, we consider some details about vertex characteristics along with network structure as the main two aspects to be taken into account by the analyst when the goal consists in characterizing the topology of a network. We make such a distinction because the former describes specific node attributes, whereas the latter characterizes global network attributes.

\subsection{Vertex characteristics}

Frequently, the first step to characterize a network consists in describing its vertices. The degree of a vertex refers to the number of edges connected to that vertex; this quantity allows us to identify the most highly connected vertices in the network. The degree distribution in most real-world networks is highly right-skewed, and therefore very unlike the random graph case (see Section \ref{sec_statistical_models_for_networks}); indeed, many of them follow power laws in their tails (i.e., $p_k\propto k^{-\gamma}$ where $p_k$ is the fraction of vertices with degree $k$, and $\gamma$ is some exponent greater than zero). From a structural perspective, it is useful to look at the average neighbor degree (two vertices are referred to as neighbors if they are joined by an edge) versus the vertex degree in order to investigate how vertices of different degrees are linked to each other.

Vertex centrality measures allow us to characterize the relative importance of an actor in the network. Obviously, the definition of these measures depends on the underling notion of \quo{importance}. For instance, closeness centrality measures suggest that a vertex is important if it is close to many other vertices, while betweenness centrality measures label a vertex as important if it is between many other pair of vertices. Centrality measures are usually based on the geodesic distance, i.e., the length of the shortest path between vertices. Many other centrality measures have been proposed over the years; see for example \citet[Ch. 4]{kolaczyk-2009} for a review.

In addition to describing vertices' characteristics, it is also very important to characterize the network's structure as a whole. In what follows, we review some measures about this regard.

\subsection{Network structure}

Two fundamental aspects of the structure of a network are cohesion and connectivity. Of course, there are several ways to assess cohesiveness attributes. One way to do so simply consists in establishing whether or not the network is connected (i.e, every vertex is reachable from every other vertex) or complete (i.e, every vertex is joined to every other vertex), and enumerating pre-specified subgraphs of interest such as dyads (pairs), triads (triples) or cliques (undirected graph such that every two distinct vertices in the clique are adjacent).

There are also several measures specifically designed to describe connectivity in a network. For instance, the density of a network, defined as the frequency of realized edges relative to the number of potential edges, measures how close the network is to being complete. In addition, the clustering coefficient or transitivity, defined as the relative frequency of connected triples to triangles (three nodes connected to each other by three edges), measures the density of triangles in the network and therefore its transitivity. Density and clustering in the immediate neighborhood of a vertex are also possible. Another way of examining connectivity is related to the impact that vertex removal might have on the existence of paths between pairs of vertices; this notion is commonly known as resilience. In many real networks, only a few percent of hight degree vertices need be removed before essentially all communication through the network is destroyed.

In most kinds of networks there are different types of vertices according to certain attributes. Selective linking among vertices according to these characteristics is usually called homophily or assortative mixing. Homophily provides an explanation to patterns often seen in social networks, such as transitivity (\quo{a friend of a friend is a friend}), balance (\quo{the enemy of my friend is an enemy}), and the existence of cohesive subgroups of nodes \cite[p. 1]{hoff-2008}. Measures that aim to quantify the extent of homophily are called assortative coefficients and essentially are variations of a regular correlation coefficient. One common use of assortative coefficients consists in summarizing the degree correlation of adjacent vertices.

As an extreme case of homophily, it is common to find subsets of actors that demonstrate cohesive patterns with respect to the underlying relational framework. Such groups of vertices have a high density of edges within them, with a lower density of edges between groups. Networks evidencing this behavior are said to have a community structure. In that regard, hierarchical clustering and spectral partitioning are two classical methods often used to detect network communities in the absence of external information. Specifically, hierarchical clustering methods aim to algorithmically optimize a similarity measure in order to detect vertices in the same communities (e.g., two vertices can be considered as similar if they have the same neighbors). On the other hand, spectral partitioning methods attempt to discover communities by iteratively using the eigen-decomposition of the graph Laplacean. Development of procedures for community detection is a highly active area of research. There are numerous reviews available; see \citet{fortunato-2010}, for example.

\section{Modeling}\label{sec_statistical_models_for_networks}

Generally speaking, a statistical network model is a probability distribution on a sociomatrix $\Y$ indexed by an unknown parameter $\tev\in\Theta$, $p(\Y\mid\tev)$. Rather than visualizing and describing topological characteristics of the network, statistical models aim to study essential aspects of the stochastic mechanism under which a given network might have arisen. Indeed, statistical network models allow us to test for the significance of predefined features in the network, assess associations between node/edge attributes and the network structure, and impute missing observations. In contrast to deterministic and algorithmic models, statistical models are also useful to quantify the uncertainty related to the unknowns in the model (e.g, parameter estimates, predictions, and missing data imputations).

It is very important to emphasize that the nature of a network itself leads to dependencies between actors, and also, between ties; for instance, reciprocity and clustering are clear manifestations of dependence in network data. It is indispensable to take such dependencies into account if we want to formulate reasonable statistical models. A concise discussion of relevant models for cross-sectional (also called static) networks is presented below. An extensive treatment of these topics can be found for example in \citet{goldenberg-2010}, \citet{snijders-2011}, and \citet{crane-2018-probabilistic}.

\subsection{Random graph models}

Statistical models for networks have now over 50 years of history. The random graph model \citep{gilbert-1959,erdos-1959,erdos-1960,erdos-1961} was one of the first models for networks discussed in the literature. Under this model, an edge between any pair of nodes is added to the graph independently with some fixed probability $\theta$.  For example, the probability of an undirected, binary network under this model is given by
\begin{equation*}
p(\Y\mid\theta) = \prod_{i<i'}\theta^{y_{i,i'}}(1-\theta)^{1-y_{i,i'}}.
\end{equation*}
Random graphs tend to be sparse with small diameter (value of the longest geodesic distance), low clustering, and an unrealistic degree distribution. Hence, most real-world networks are rarely a plausible realization of a random graph. In spite of such unrealistic behavior, random graph models are commonly used in defining null classes of networks against which to assess the significance of structural characteristics found in an observed network \citep[Sec. 5.5, for example]{kolaczyk2020statistical}. \citet{bollobas-1998} offers an extensive treatment of random graph models.

\subsection{Generalized random graph models}

Motivated by real-world network attributes, generalized random graph models arose as an extension of the original random graph aiming to mimic such attributes through the inclusion of simple mechanisms. For instance, configuration models \citep{bender-1978} generate random networks with a pre-specified degree distribution. A shortcoming of such a model is that it fails at capturing homophily and clustering, which are features frequently observed in social networks. On the other hand, small-world models \citep{watts-1998,newman-1999} produce high levels of clustering with small average distances, but generate unrealistic degree distributions. In addition, both the configuration model and the small-world model work with a fixed number of nodes and thus cannot be used to model network growth (phenomenon in which the number of nodes in the network increases over time). On the other hand, preferential attachment models \citep{barabasi-1999}, designed to account for network growth and preferential attachment (\quo{the rich get richer} effect), yield networks with degree distributions that tend to a power law. Nevertheless, this model still shares the tendency towards low clustering. As a consequence, neither the configuration model, the small-world model, nor the configuration model should be viewed as fully realistic models for networks. \citet{chung-2006} is a classical reference on generalized random graphs models.

\subsection{Exponential random graph models}

Beyond generalized random graphs models, \citet{frank-1986} introduced the so-called exponential random graphs models (ERGMs), also known as $p^*$ models \citep{wasserman-1996}, attempting to built more realistic models to address the foregoing transitivity issue. Specifically, ERGMs can be written as
\begin{equation}\label{eq_ERGM}
p(\Y\mid\X,\tev) = \frac1{\kappa(\tev)}\,\ex{\sum_{k=1}^K \theta_k S_k(\Y,\X)}
\end{equation}
where $\X$ is an array of predictors $\xv_{i,i'}=(x_{i,i',1},\ldots,x_{i,i',P})$ specific to each dyad $(i,i')$,
each $S_k(\Y,\X)$ is either a network statistic or a function of edge and vertex attributes, $\tev = (\theta_1,\ldots,\theta_K)$ is a $K$-dimensional vector of unknown parameters, and $\kappa(\tev)$ is a normalizing constant. Examples of network statistics include counts of $k$-stars ($k+1$ nodes with one node being linked to the other $k$) and triangles. ERGMs are appealing models for networks since the form of \eqref{eq_ERGM} explicitly tie parameters to sufficient statistics, yielding an attractive interpretation. Furthermore, ERGMs can be constructed to match beliefs on important structural features of the data.

Even though ERGMs have a natural appeal, they are computationally challenging because the normalizing constant $\kappa(\tev)$ is generally unknown and intractable in all but the simplest cases. An additional shortcoming is that ERGMs tend to degenerate, i.e., the model places disproportionate probability mass on only a few of the possible graph configurations. Also, ERGMs implicitly assume that the network is observed for the whole population of interest, and therefore, they are not well suited to make predictions on unobserved dyads. Finally, a recognized limitation of ERGMs is that they are weak at capturing local features of networks and as a consequence may lead to poor model fitting in real-world networks \citep{snijders-2002,handcock-2003}.  \citet{frank-1986} also proposed models with Markov structure that provide forms of dyad dependence (homogeneous monadic Markov models). A detailed review of ERGMs can be found in \citet{robins-2007} and \citet{lusher-2012}.

\subsection{Social space models}\label{sec_social_space_models}

The use of random effects in the context of generalized linear models is a popular alternative to model networks. Specifically, consider a model in which the $y_{i,i'}$s are conditionally independent with probabilities of interaction
\begin{equation}\label{eq_model_1}
\pr{y_{i,i'}=1\mid \bev, \gamma_{i,i'},\xv_{i,i'} } = g^{-1}(\xv^\trans_{i,i'}\bev + \gamma_{i,i'}),\quad i<i',
\end{equation}
where $\boldsymbol{\beta}=(\be_1,\ldots,\be_P)$ is an (unknown) vector of fixed effects, $\xv^\trans_{i,i'}\bev = \sum_{p=1}^P \beta_p\,x_{i,i',P}$ is a linear predictor representing patterns in the data related to known covariates $\xv_{i,i'}$, $\gamma_{i,i'}$ is an unobserved specific random effect, representing any additional patterns in the data unrelated to those of the predictors, and $g(\cdot)$ is a (known) link function.

Following results in \citet{aldous-1985} and \citet{hoover-1982}, see also \citet{hoff-2008} for details, it can be shown that if the matrix of random effects $[\gamma_{i,i'}]$ is jointly exchangeable, there exists a symmetric function $\alpha(\cdot,\cdot)$ such that $\gamma_{i,i'}=\al(\uv_i,\uv_{i'})$, where $\uv_1,\ldots,\uv_I$ is a sequence of independent latent random variables (vectors). The impact of such latent variables on \eqref{eq_model_1} is largely dictated by the form of $\alpha(\cdot,\cdot)$. Therefore, it is mainly through $\alpha(\cdot,\cdot)$ that we are able to capture relevant features of relational data.

A number of potential formulations for $\alpha(\cdot,\cdot)$ have been explored in the literature to date; for instance, see \citet{nowicki-2001}, \citet{hoff-2002}, \citet{schweinberger-2003}, \citet{hoff-2005}, \citet{handcock-2007}, \citet{linkletter2007spatial}, \citet{hoff-2008}, \citet{krivitsky-2008}, \citet{hoff-2009}, \citet{krivitsky-2009}, \citet{li-2011}, \citet{raftery-2012}, and \citet{minhas2019inferential}. Some of these approaches are discussed bellow (see also table \ref{tab_summary_models}).  Other important approaches in a multilayer setting include \citet{salter2017latent}, \citet{durante2018bayesian}, y \citet{wang2019common}. See also Section \ref{sec_discussion} for a discussion.

\subsubsection{Class models}

\citet{nowicki-2001} assume that each actor $i$ belongs to an unobserved latent class $u_i\in\{1,\ldots,K\}$, and a probability distribution describes the relationships between each pair of classes. Here, the latent effects are specified as $\al(u_i,u_{i'})=\theta_{\phi(u_i,u_{i'})}$, for a symmetric $K\times K$ matrix $\Theta=[\theta_{k,\ell}]$ of real entries $\theta_{k,\ell}$ such that $0 < \theta_{k,\ell} < 1$, with $\phi(u,v)=(\min\{u,v\},\max\{u,v\})$. Latent class models, also known as stochastic block models (SBMs), effectively capture stochastic equivalence (pattern in which nodes can be divided into groups such that members of the same group have similar patterns of relationships). However, models based on distinct clusters may not fit well when many actors fall between clusters \cite{hoff-2002}. Recent extensions of this approach are given in \citet{kemp-2006}, \citet{xu-2006}, and \citet{airoldi-2009}.

\subsubsection{Distance models}

\citet{hoff-2002} assume that each actor $i$ has an unknown position $\uv_i\in\reals^K$ in an Euclidean social space (space of unobserved latent characteristics that represent potential transitive tendencies in network relations), and that the probability of an edge between two actors may increase as the latent characteristics of the individuals become more similar, i.e., when the actors become closer in the social space. To this end, the latent effects are specified as $\alpha(\uv_i,\uv_{i'})=-\|\uv_i-\uv_{i'}\|$, where $\|\cdot\|$ denotes the Euclidean norm. Latent structures based on distances naturally induce homophily (pattern in which the relationships between nodes with similar characteristics are stronger than those between nodes with different characteristics), which is a main feature frequently seen in real social networks. Also, modeling positions as belonging to a low-dimensional Euclidean space provides a model-based alternative of data reduction to graphically represent social network data. Even though latent distance models inherently account for reciprocity and transitivity, they may not be appropriate for networks exhibiting hight levels of clustering.

\subsubsection{Projection models}

\citet{hoff-2002}, in the same context of latent distance models where $\uv_i\in\reals^K$, propose that the probability of an edge between two actors may increase as the overture of the angle formed by the corresponding latent positions becomes wider. Specifically, actors $i$ and $i'$ are prone to having a tie if the angle between them is small ($\uv_i^\trans\uv_{i'} > 0$), neutral to having ties if the angle is a right angle ($\uv_i^\trans\uv_{i'} = 0$), and averse to ties if the angle is obtuse ($\uv_i^\trans\uv_{i'} < 0$). The latent effects are specified as $\al(\uv_i,\uv_{i'})=\uv_i^\trans\uv_{i'}/\|\uv_{i'}\|$, which corresponds to the signed magnitude of the projection of $\uv_i$ in the direction of $\uv_{i'}$. Such a quantity can be thought of as the extent to which $i$ and $i'$ share characteristics, multiplied by the activity level of $i$.

\subsubsection{Bilinear models}

\citet{hoff-2005}, considering again a $K$-dimensional social space, assumes that interaction probabilities rely on symmetric multiplicative interaction effects. Such interaction for a dyad $(i,i')$ is expressed in terms of a bilinear effect, i.e., the inner product between unobserved characteristic vectors specific to actors $i$ and $i'$. Hence, the latent effects are specified as $\al(\uv_i,\uv_{i'}) = \uv_i^\trans\uv_{i'}$. According to \citet{hoff-2008}, bilinear models are able to generalize distance models (but not class models) and reproduce different degrees of balance and clusterability.

\subsubsection{Spatial process models}

\citet{linkletter2007spatial}, extrapolating ideas from Hoff's latent distance model to a covariate space, assumes that pairwise connections are conditionally independent given a latent spatial process evaluated at observed covariates. Thus, the probability of an edge between actors $i$ and $i'$ depends on a relative difference between observed covariates $\xv_i$ and $\xv_{i'}$, through latent effects expressed as $-\|z(\xv_i)-z(\xv_{i'})\|$, where $z(\cdot)$ is a latent real-valued function. Note that in this context, the $z(\xv_i)$ are actually unobserved, and the covariates $\xv_i$ represent attributes measured to learn about social relations.

\subsubsection{Cluster models}

\citet{handcock-2007}, \citet{krivitsky-2008}, and \citet{krivitsky-2009} generalize Hoff's latent distance model in an effort to recreate a model that allow the practitioner to model both transitivity and homophily, and simultaneously find clusters of actors in a model-based fashion when the number of groups in the data is known. As in \citet{hoff-2002}, the latent effects are given by $\alpha(\uv_i,\uv_{i'})=-\|\uv_i-\uv_{i'}\|$, except that now, actors' positions are drawn from a finite spherical multivariate normal mixture. Thus, the position of each actor is drawn from one of $G$ groups, where each group is centered on a different mean vector and dispersed with a different spherical covariance matrix, which allow latent positions form cluster of actors within the latent space. Note that the model of Hoff's distance model is essentially the case with $G = 1$.

\subsubsection{Eigen models}

\citet{hoff-2008} and \citet{hoff-2009}, based on the principles of eigen-analysis, assume that the relationship between two nodes as the weighted inner-product of node-specific vectors of latent characteristics $\uv_i\in\reals^K$. Here, the latent effects have the form $\al(\uv_i,\uv_{i'})=\uv_i^\trans\LAM\uv_{i'}$, where $\LAM$ is a $K\times K$ diagonal matrix. These models, also known as eigenmodels, generalize latent class and latent distance models in the sense that they can compactly represent the same network features, but not vice versa. As a result, eigenmodels can represent both positive or negative homophily in varying degrees, and stochastically equivalent nodes may or may not have strong relationships with one another \citep{hoff-2008}.

\begin{table}[!h]
	\centering
	\begin{tabular}{lll}  
		\hline
		Model &  Latent effects & Latent space \\ 
		\hline
		Class      & $\al(u_i,u_{i'})=\theta_{\phi(u_i,u_{i'})}$  & $u_i\in\{1,\ldots,K\}$ \\ 
		Distance   & $\alpha(\uv_i,\uv_{i'})=-\|\uv_i-\uv_{i'}\|$ &  $\uv_i\in\reals^K$    \\
		Projection & $\al(\uv_i,\uv_{i'})=\uv_i^\trans\uv_{i'}/\|\uv_{i'}\|$ & $\uv_i\in\reals^K$ \\
		Bilinear   & $\al(\uv_i,\uv_{i'}) = \uv_i^\trans\uv_{i'}$ & $\uv_i\in\reals^K$ \\
		Spatial process & $\al(\xv_i,\xv_{i'}) = -\|z(\xv_i)-z(\xv_{i'})\|$ & $\xv_i\in\mathcal{X}^P$ \\
		Cluster & $\alpha(\uv_i,\uv_{i'})=-\|\uv_i-\uv_{i'}\|$ &  $\uv_i\in\reals^K$ \\
		Eigen  & $\al(\uv_i,\uv_{i'})=\uv_i^\trans\LAM\uv_{i'}$ & $\uv_i\in\reals^K$ \\
		\hline
	\end{tabular}
	\caption{{\footnotesize Summary of latent space models.}}\label{tab_summary_models} 
\end{table}

\section{Computation}\label{sec_computation}

For a given $K$ the posterior distribution of the parameters can be explored using Markov chain Monte Carlo \citep[MCMC]{gamerman2006markov} algorithms in which the posterior distribution is approximated using dependent but approximately identically distributed samples 
$\UPS^{(1)},\ldots,\UPS^{(B)}$, with $\UPS= \left( \uv_1,\ldots,\uv_I, \phiv \right)$,
where $\phiv$ has as elements the rest of the model parameters. Point and interval estimates can be approximated from the empirical distributions. Details about MCMC algorithms implemented here can be found in Appendix \ref{app_mcmc}.

\section{Illustrations}\label{sec_illustrations}

In what follows, we present some examples in which we fully implement some of the latent space models described in Section \ref{sec_social_space_models}. We illustrate the characteristics of these models by analyzing popular datasets in the network literature. Special attention is given to latent class, distance, and eigen models.

\subsection{Florentine families dataset}

\begin{figure}[h!]
	\centering
	\subfigure[Adjacency matrix.]{\includegraphics[width=.47\linewidth]{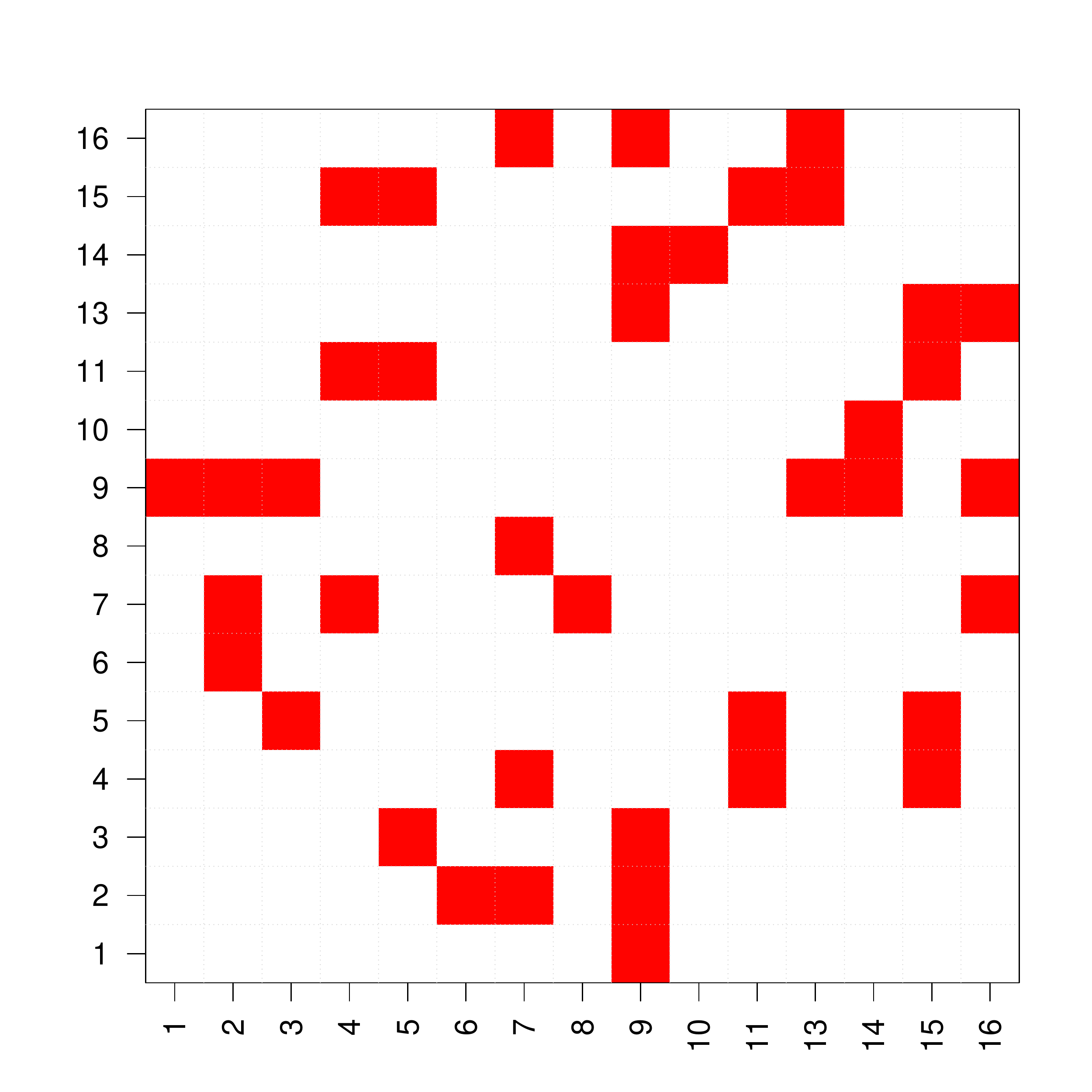}}
	\subfigure[Interaction probabilities.]{\includegraphics[width=.47\linewidth]{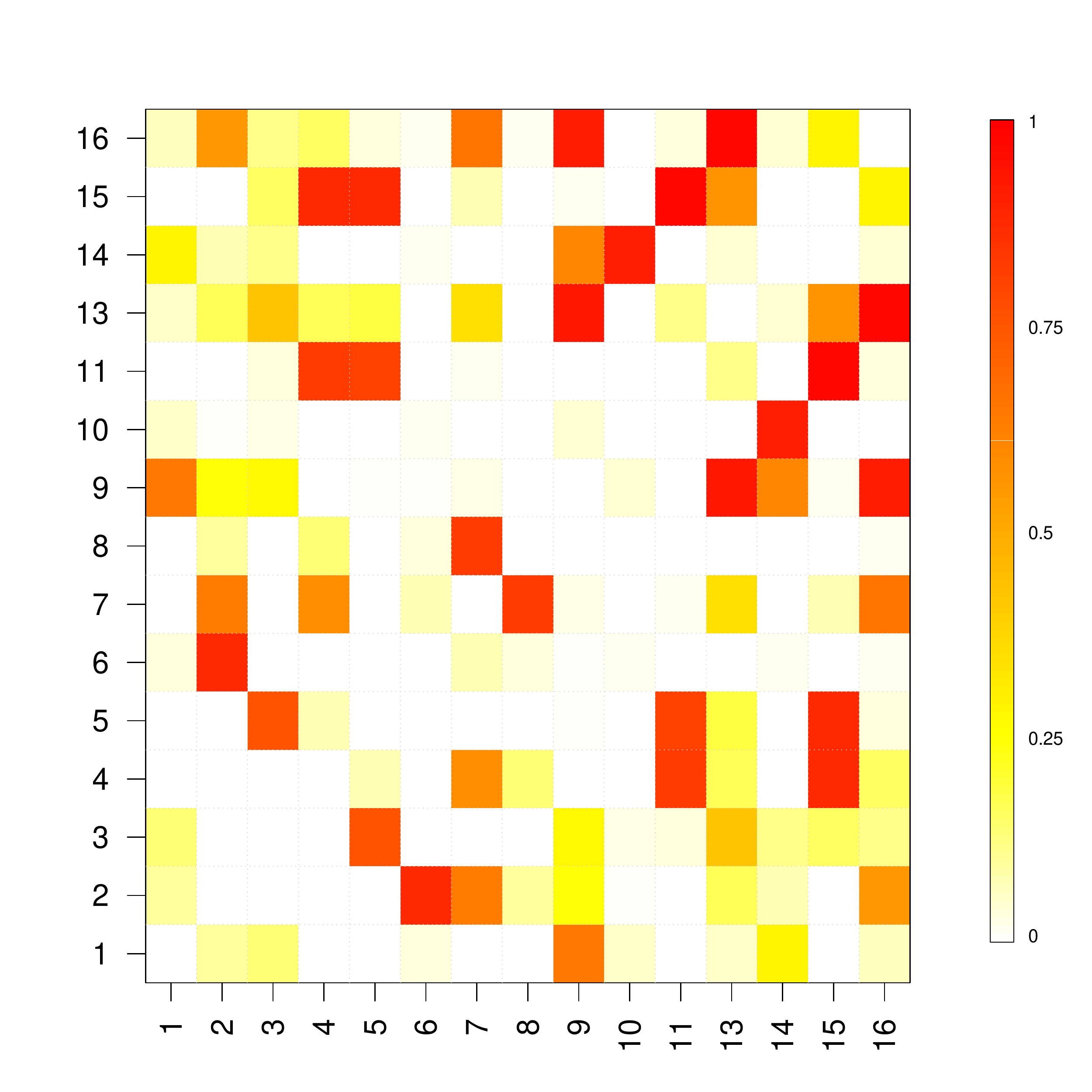}}
	\subfigure[Latent positions.]{\includegraphics[width=.47\linewidth]{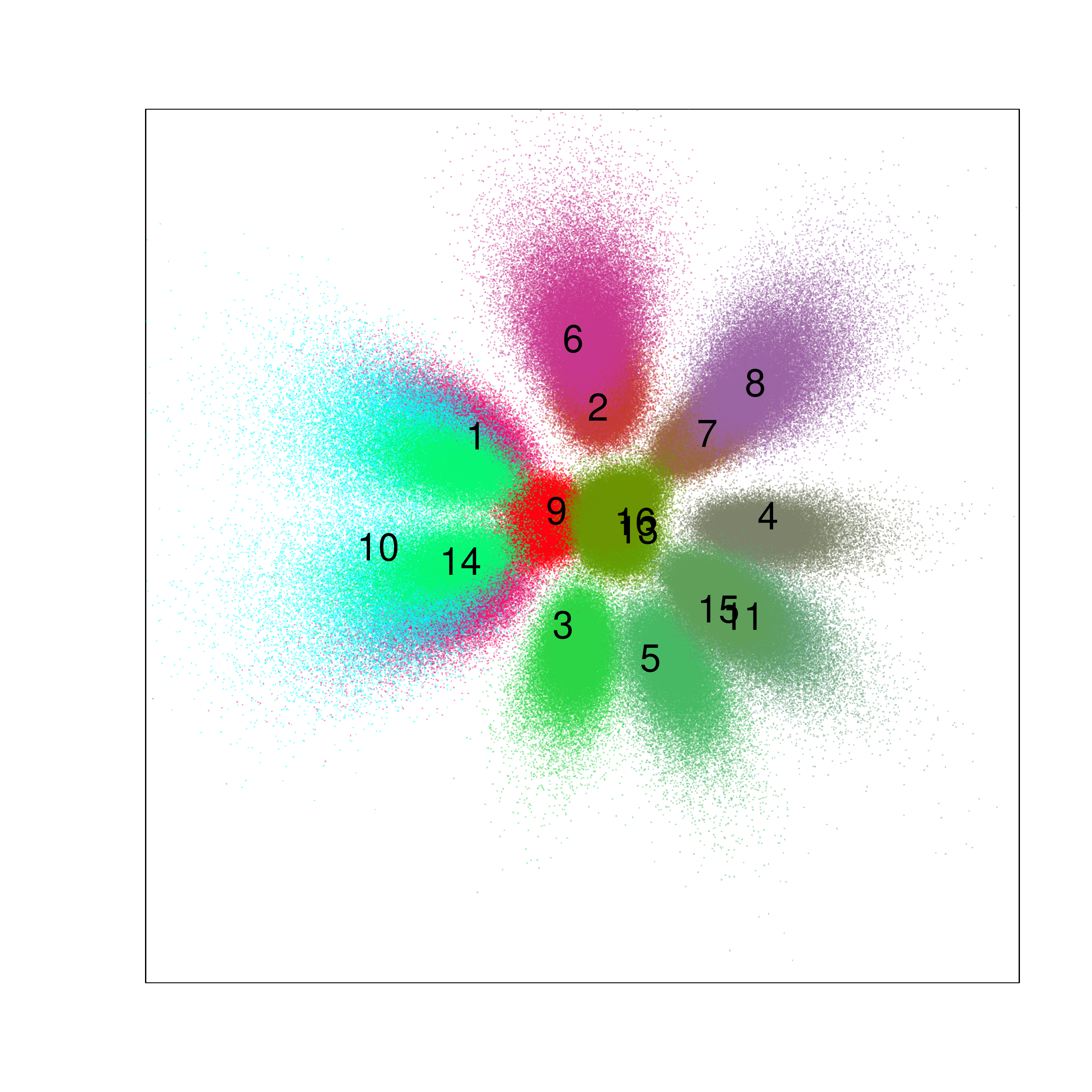}}
	\caption{\footnotesize{Florentine families dataset.}}
	\label{fig_padgm}
\end{figure}

Here, we illustrate a fully Bayesian implementation of the distance model by reproducing the analysis of the florentine families dataset given in \citet[Section 4.2]{hoff-2002}. The system is composed of $I = 15$ prominent families, for which $y_{i,i'} = 1$ between families $i$ and $i'$ if there is at least one marriage between them. We considered this as an undirected relation, whose corresponding adjacency matrix $\Y$ is displayed in Panel (a) of Figure \ref{fig_padgm}.

We consider a latent space with $K=2$ dimensions, which also will help us to demonstrate the graphical capabilities of the model. Indeed, setting $K=2$ simplifies visualization and interpretation, and is therefore particularly useful when the main goal of the analysis is to provide a description of the social relationships. Following Section \ref{sec_social_space_models}, we implement a model of the form,
$$
y_{i,i'}\mid\zeta,\uv_i,\uv_{i'} \simind \Ber\left(\expit\left(\zeta - \|\uv_i-\uv_{i'}\|\right)\right)\,,
$$
where $\expit(x)=1/(1+e^{-x})$ is the inverse of the logit function, $\zeta$ is a fixed effect representing the average propensity of observing an edge between two given actors, and $\uv_1,\ldots,\uv_I$ are unobserved positions in $\reals^2$. In order to proceed with a fully Bayesian analysis and make inference about the model parameters, we must specify prior distributions for $\zeta$ and each $\uv_i$. A standard prior choice that seems to work well in practice is $\zeta\mid\ome^2 \sim \Nor(0,\ome^2)$ and $\uv_i\mid\sig^2\simiid \Nor(\zerov,\sig^2\,\I)$, where $\I$ denotes the identity matrix. We complete the formulation of the model by letting $\ome^2 \sim \IGamd(a_\ome,b_\ome)$ and $\sig^2\sim\IGamd(a_\sig,b_\sig)$. Sensible elicitation of the hyperparameters $a_\ome$, $b_\ome$, $a_\sig$, and $b_\sig$ is fundamental to ensure appropriate model performance. To this end, we set $a_\ome = 2$ and $b_\ome = 100$, which places a diffuse prior distribution for $\zeta$. Similarly, we mimic a heuristic given in \citet[Sec. 2.4]{krivitsky-2008} by setting $a_\sig$ and $b_\sig$ in such a way that a priori $\sig^2$ is vaguely concentrated (e.g., $\coefvar{\sig^2} = 1$) around $\expec{\sig^2}= \tfrac{\pi}{ \Gamma(2)}\, I^{2/K}$, i.e., the volume of a $2$-dimensional Euclidean ball of radius $I^{1/K}$.

Markov chain Monte Carlo (MCMC) algorithms can be used to explore the posterior distribution $p(\zeta,\U,\ome^2,\sig^2\mid\Y)$, where $\U = [\uv_1,\ldots,\uv_I]^\trans$ is a $I\times K$ matrix storing the latent positions by rows. By means of the MCMC procedure outlined in Section \ref{app_mcmc_distance}, we obtain $50,000$ samples of the posterior distribution after a burn-in period of $10,000$ iterations. In this case and subsequent illustrations, convergence was monitored by tracking the variability of the joint distribution of data and parameters using the multi-chain procedure discussed in \cite{GeRu92}.

Notice that an inherent difficulty estimating $\U$ is that any rotation, reflection or translation of $\U$ produce the same likelihood value. Indeed,
for any $K\times K$ orthogonal matrix $\Q$, the likelihood associated with the reparametrization $\tilde{\uv}_i = \Q\uv_i$ is independent of $\Q$, since $\|\uv_i -\uv_{i'}\| = \|\tilde\uv_i -\tilde\uv_{i'}\|$. To address this issue, we restrict our attention to the Procrustean transformation of $\U$ closest to a fixed (but arbitrary!) reference configuration $\U_0$. In particular, we consider a post-processing step in which posterior samples are rotated/reflected to a shared coordinate system. Thus, for each sample $\UPS^{(b)}$, an orthogonal transformation matrix $\Q^{(b)}$ is obtained by minimizing the Procrustes distance,
\begin{align}\label{eq_proc}
\tilde{\Q}^{(b)} = \argmin_{\Q \in \mathcal{S}^{K}} \tr\left\{ \left( \U_0-\U^{(b)}\Q \right)^\trans \left( \U_0-\U^{(b)}\Q \right) \right\}\,,
\end{align}
where $\mathcal{S}^{K}$ denotes the set of $K \times K$ orthogonal matrices. The minimization problem in \eqref{eq_proc} can be easily solved using singular value decompositions \citep[Section 20.2, for example]{borg-2005}.  Once the matrices $\tilde{\Q}^{(1)}, \ldots, \tilde{\Q}^{(B)}$ have been obtained, posterior inference for the latent positions are based on the transformed coordinates $\tilde{\uv}_{i}^{(b)} = \tilde{\Q}^{(b)} \uv_{i}^{(b)}$. In this case, we let $\U_0$ be the first value of $\U$ after the burn-in period of the chain. We plot the latent positions for each saved scan along with the corresponding point estimates for every family as shown in Panel (c) of Figure \ref{fig_padgm}. Actors 14 and 10 are above or below actor 1 for any particular sample; the observed overlap of these actors is due to the bimodality of the posterior distribution.

Finally, we check the posterior means of the interaction probabilities,
$$\expec{\expit\left(\zeta - \|\uv_i-\uv_{i'}\|\right)\mid\Y} \approx \frac{1}{B}\sum_{b=1}^B \expit\left(\zeta^{(b)} - \|\uv_i^{(b)}-\uv_{i'}^{(b)}\|\right) \,,$$
to examine the in-sample fit of the model. Panel (b) of Figure \ref{fig_padgm} suggests that these posterior estimates are consistent with the adjacency matrix $\Y$ plotted in Panel (a), since we see high posterior probabilities where connections are observed.

\subsection{Village dataset}

\begin{figure}[H]
	\centering
	\subfigure[Adjacency matrix.]{\includegraphics[width=.44\linewidth]{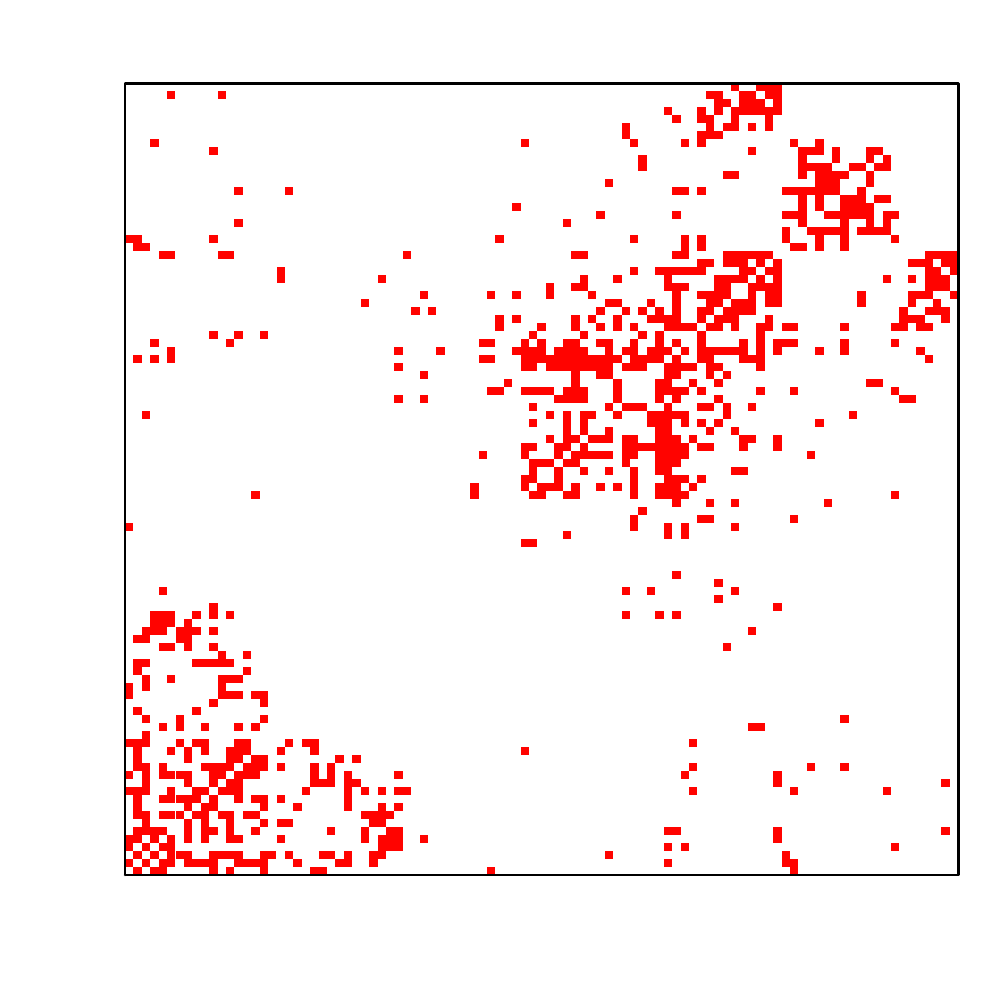}}
	\subfigure[Interacition probabilities.]{\includegraphics[width=.44\linewidth]{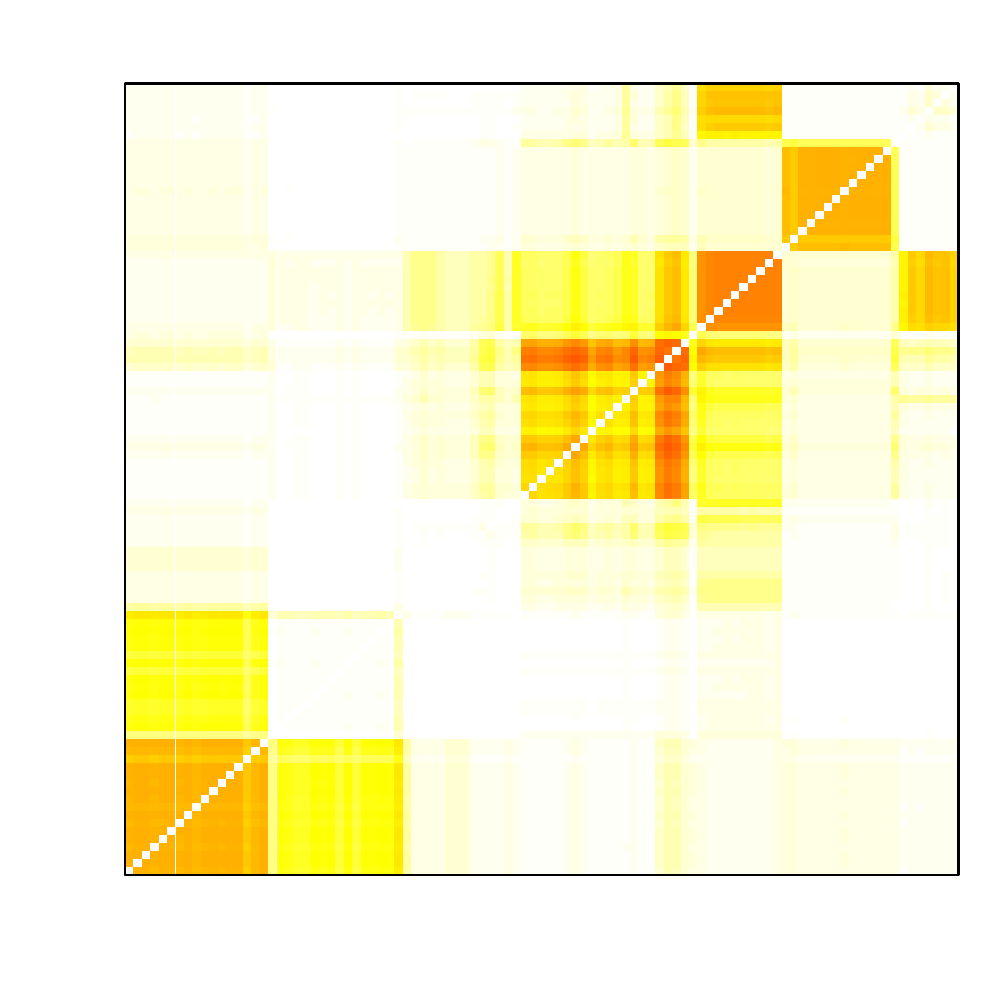}}
	\includegraphics[width=0.65\linewidth]{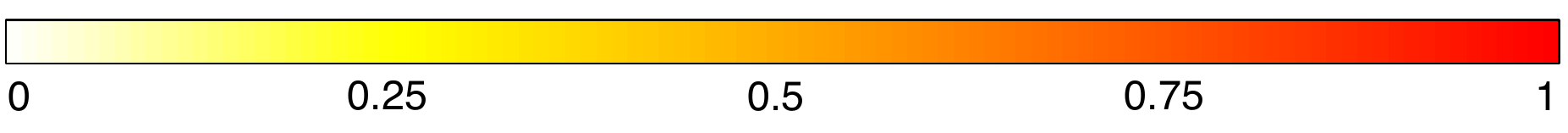}
	\subfigure[Communities point estimate.]{\includegraphics[width=.44\linewidth]{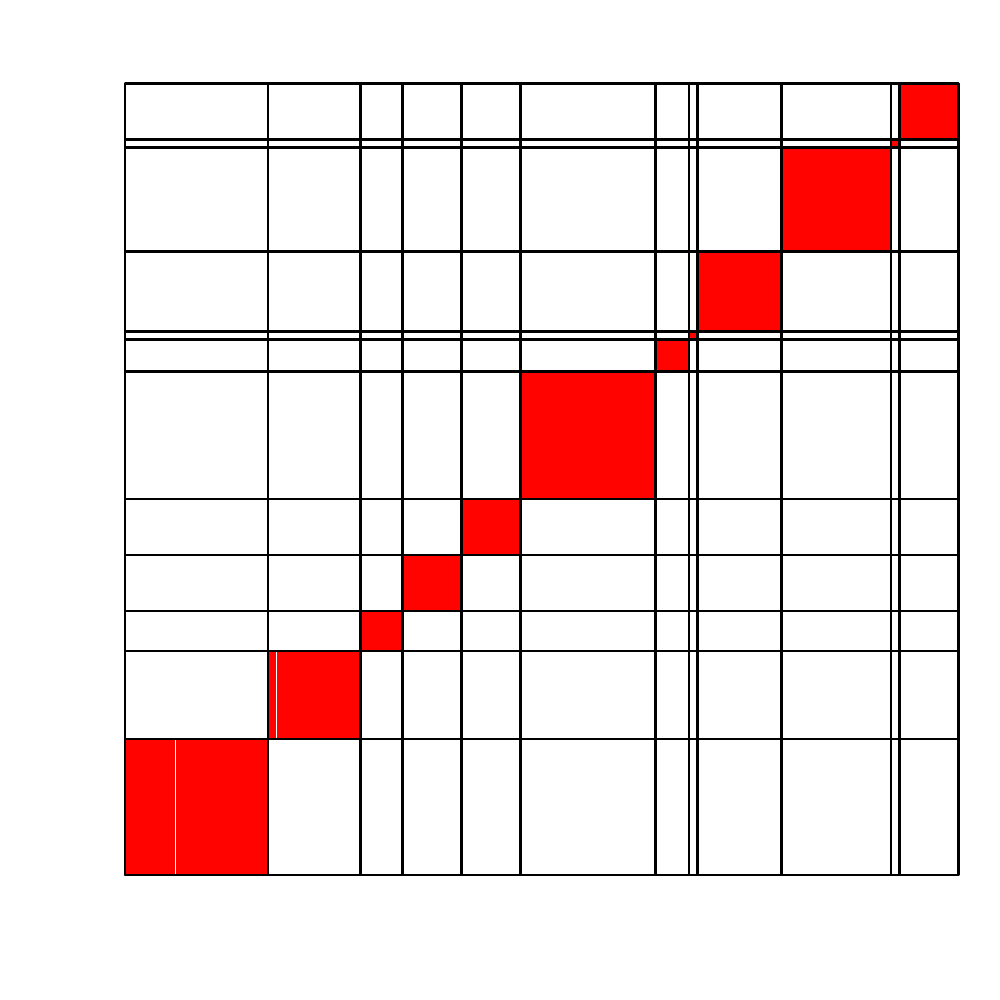}}
	\subfigure[Co-membership propabilities.]{\includegraphics[width=.44\linewidth]{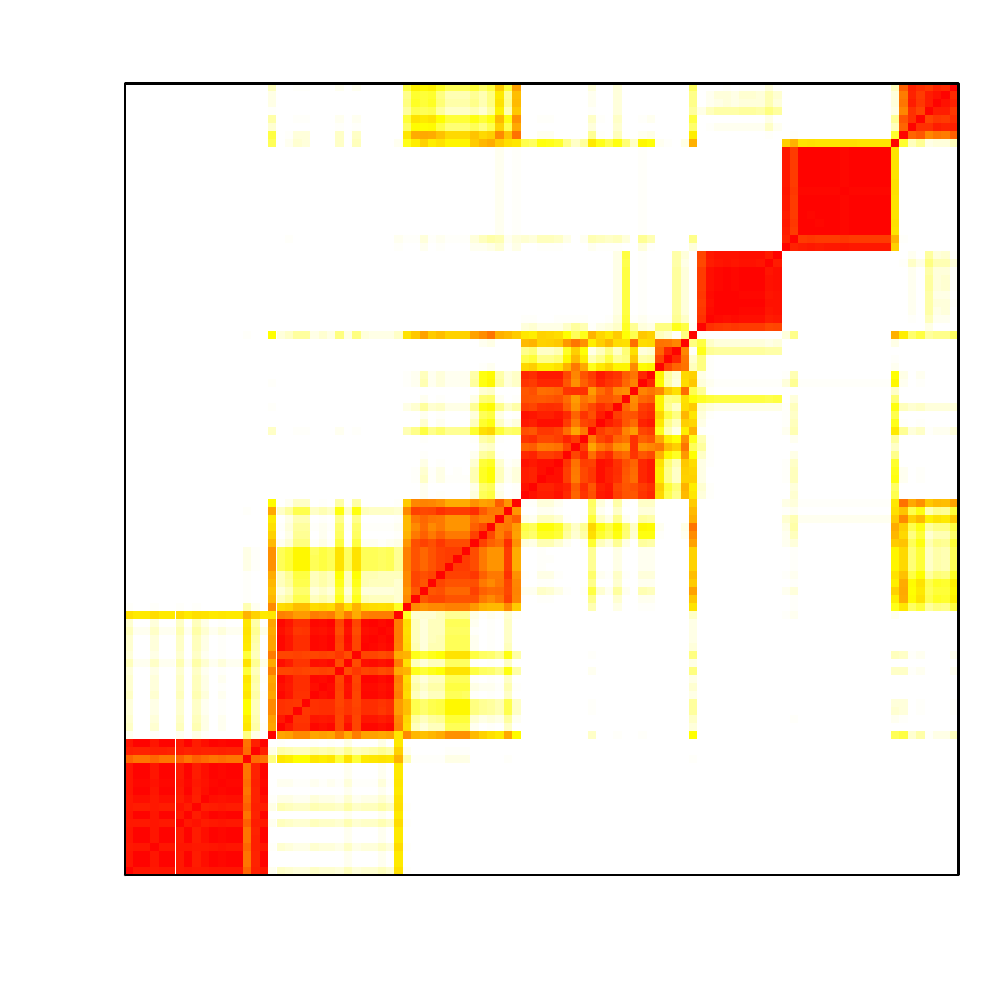}}
	\caption{\footnotesize{Village dataset.}}
	\label{fig_village}
\end{figure}

In order to provide a community detection example by means of a class model, we consider the social and familial relationships among $I = 99$
households in a specific village located in rural southern Karnataka, India \citep{salter2017latent}. For these data, $y_{i,i'} = 1$ if household $i$ and $i'$ have a social tie by being related or attending temple together, for example. The adjacency matrix $\Y$ associated with this network is depicted in Panel (a) of Figure \ref{fig_village}.

The main idea behind class models is that similar actors can be clustered together into groups known as classes or blocks. Thus, the probability of having an edge between two actor can be modeled as a function of their respective blocks,
$$
y_{i,i'}\mid u_i,u_{i'}, \{\eta_{k,\ell}\}  \simind \Ber\left( \expit\eta_{\phi(u_i,u_{i'})}  \right)\,,
$$
where $\uv = (u_1,\ldots,u_I)$ are unobserved cluster indicators taking values in $\{1,\ldots,K\}$, with $K$ the number of classes (assumed as fixed), and $\phi(a,b)=(\min\{a,b\},\max\{a,b\})$. Notice that actors $i$ and $i'	$ belong to the same class if and only if $u_i=u_{i'}$. The community parameters $\etav=\{ \eta_{k,\ell}: k,\ell = 1,\ldots,K, k \leq \ell \}$ suffer from symmetry constraints because $\Y$ is a symmetric adjacency matrix, which makes $\phi(\cdot,\cdot)$ necessary. A standard choice of prior distribution for the community parameters is achieved by letting these parameters be conditionally independent and follow a common distribution, $\eta_{k,\ell}\mid\mu,\tau^2\simiid\Nor(\zeta,\tau^2)$.

Following a standard practice in the community detection literature (\citep{nowicki-2001}, for example), it is commonly assumed that the entries of $\uv$ are exchangeable \citep[Sec. 1.2, for example]{gelman-2014-information} and follow a categorical distribution on $\{1,\ldots,K\}$, $\pr{u_i=k\mid \omega_k} = \omega_k$, $k = 1,\ldots,K$, where $\omev=(\omega_1,\ldots,\omega_K)$ is a probability vector such that $\sum_{k=1}^K\omega_k = 1$, satisfying $\omev\mid\alpha\sim\Dir\left(\tfrac{\alpha}{K},\ldots,\tfrac{\alpha}{K}\right)$. In the limit, as $K\rightarrow\infty$, this formulation has a direct connection with a Chinese restaurant process prior \citep[Sec. 3]{ishwaran2000markov}. The model is completed by placing a hyperprior distributions on $(\zeta,\tau^2,\alpha)$. A well-behaved choice is consist in independently letting $\zeta \sim \Nor(\mu_\zeta,\sigma^2_\zeta)$, $\tau^2 \sim \IGamd(a_\tau,b_\tau)$, and $\al \sim \Gamd(a_\al,b_\al)$, where $\mu_\zeta$, $\sigma^2_\zeta$, $a_\tau$, $b_\tau$, $a_\al$, and $b_\al$ are hyperparameters.

Once again, a sensible elicitation of the hyperparameters is strongly recommended to ensure appropriate model performance. To this end, we  set $\mu_\zeta = 0$, $\sig_\zeta^2=3$, $a_\tau=2$, and $b_\tau = 3$, which a priori vaguely centers the prior interaction probabilities $\expit\eta_{k,\ell}$ around 0.5 allowing a fair range of values in logit scale, and $a_\al$ and $b_\al = 1$, which places a diffuse prior distribution for $\alpha$ around 1. Choosing $K=8$ and following the MCMC algorithm provided in Section \ref{app_mcmc_class}, we obtain $50,000$ samples of the posterior distribution $p(\uv,\etav,\omev,\zeta,\tau^2,\alpha\mid\Y)$ after a burn-in period of $10,000$ iterations, in order to compute the interaction probabilities and pairwise co-membership probabilities, respectively,
$$
\expec{\expit\eta_{\phi(u_i,u_{i'})}\mid\Y} \approx \frac{1}{B}\sum_{b=1}^B \expit\eta^{(b)}_{\phi(u_i,u_{i'})}
\text{ and }
\pr{u_i=u_{i'}\mid\Y} \approx \frac{1}{B}\sum_{b=1}^B \left[ u_i^{(b)} = u_{i'}^{(b)} \right]\,,
$$
where $[\cdot]$ denotes the Iverson bracket. We are quite confident about the in-sample adecuacy of the model because the interaction probabilities shown in Panel (b) of Figure \ref{fig_village} resemble very closely the adjacency matrix $\Y$ provided in Panel (a). On the other hand, We can obtain a point estimate of the communities by taking as input the co-membership probabilities shown in Panel (d) and employing the clustering methodology proposed in \cite[Sec. 4]{lau-2007} with a relative error cost of 0.5. Panel (c) provides a visual representation of such an estimate, which exhibits 12 communities with sizes ranging from 1 to 17. Notice that the pre-specified number of communities $K$ used to fit the model does not have to coincide necessarily with the number of communities provided by point estimate of the partition.

\begin{table}[!h]
	\centering
	\begin{tabular}{lccccc}  
		\hline
		Acronym &  N\underline{o} of actors & N\underline{o} of edges & Dens. & Trans. & Assor. \\ 
		\hline
		\textsf{zach}       & 34  & 78    & 0.139 & 0.256 & -0.476 \\ 
		\textsf{bktec}      & 34  & 175   & 0.312 & 0.476 &  0.015 \\ 
		\textsf{foot}       & 35  & 118   & 0.198 & 0.329 & -0.176 \\ 
		\textsf{lazega}     & 36  & 115   & 0.183 & 0.389 & -0.168 \\ 
		\textsf{hitech}     & 36  & 91    & 0.144 & 0.372 & -0.087 \\ 
		\textsf{kaptail}    & 39  & 158   & 0.213 & 0.385 & -0.183 \\ 
		\textsf{bkham}      & 44  & 153   & 0.162 & 0.497 & -0.391 \\ 
		\textsf{dol}        & 62  & 159   & 0.084 & 0.309 & -0.044 \\ 
		\textsf{glossgt}    & 72  & 118   & 0.046 & 0.184 & -0.158 \\ 
		\textsf{lesmis}     & 77  & 254   & 0.087 & 0.499 & -0.165 \\ 
		\textsf{salter}     & 99  & 473   & 0.098 & 0.335 & -0.064 \\ 
		\textsf{polbooks}   & 105 & 441   & 0.081 & 0.348 & -0.128 \\ 
		\textsf{adjnoun}    & 112 & 425   & 0.068 & 0.157 & -0.129 \\ 
		\textsf{football}   & 115 & 613   & 0.094 & 0.407 &  0.162 \\ 
		\textsf{nine}       & 130 & 160   & 0.019 & 0.163 & -0.197 \\ 
		\textsf{gen}        & 158 & 408   & 0.033 & 0.078 & -0.254 \\ 
		\textsf{fblog}      & 192 & 1,431 & 0.078 & 0.386 &  0.012 \\ 
		\textsf{jazz}       & 198 & 2,742 & 0.141 & 0.520 &  0.020 \\ 
		\textsf{partner}    & 219 & 630   & 0.026 & 0.107 & -0.217 \\ 
		\textsf{indus}      & 219 & 630   & 0.026 & 0.107 & -0.217 \\ 
		\textsf{science}    & 379 & 914   & 0.013 & 0.431 & -0.082 \\ 
		\hline
	\end{tabular}
	\caption{{\footnotesize Network datasets for which a series of cross-validation experiments are performed using distance, class, and eigen models. Dens., Trans., and Assor. stand for density, transitivity, and assortativity, respectively.}}\label{tab_datasets} 
\end{table}

\subsection{Predictive accuracy and goodness-of-fit}

\begin{figure}[!htb]
	\centering
	\subfigure[\textsf{jazz}.]      {\includegraphics[scale=0.45]{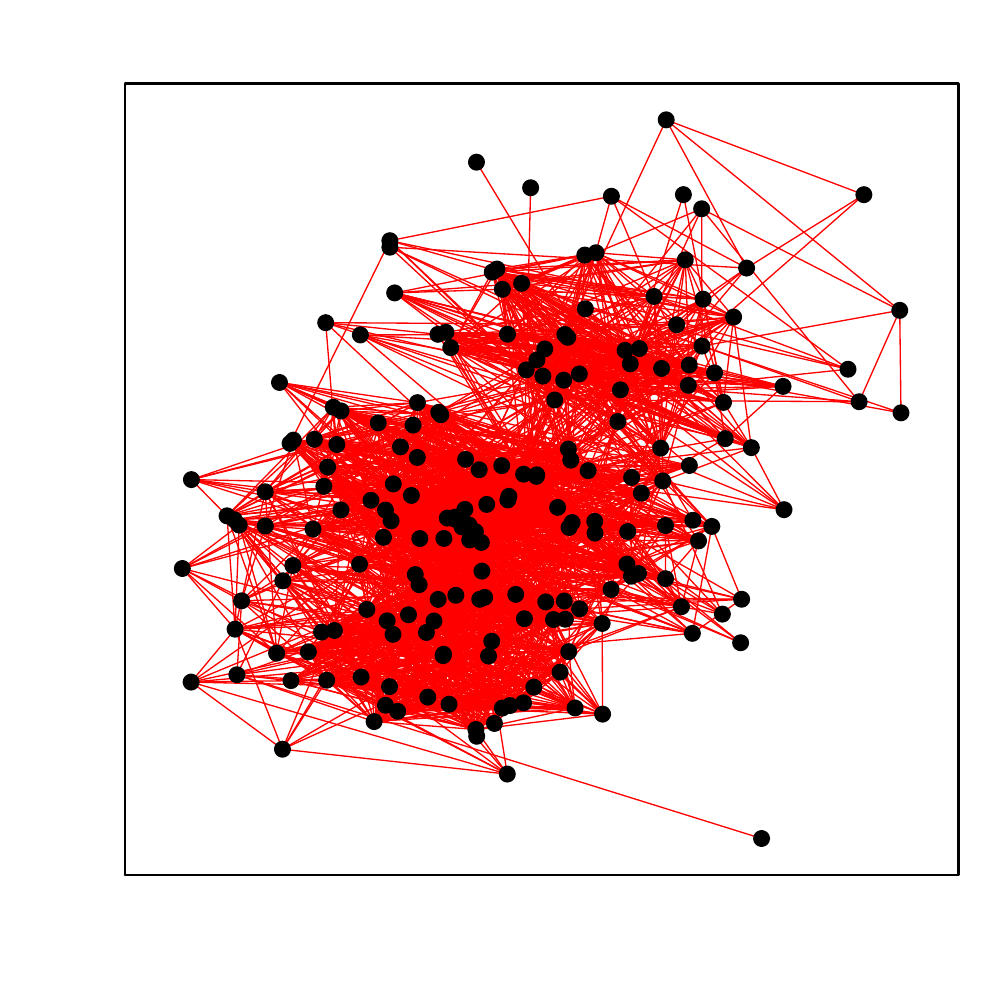}}
	\subfigure[\textsf{gen}.]       {\includegraphics[scale=0.45]{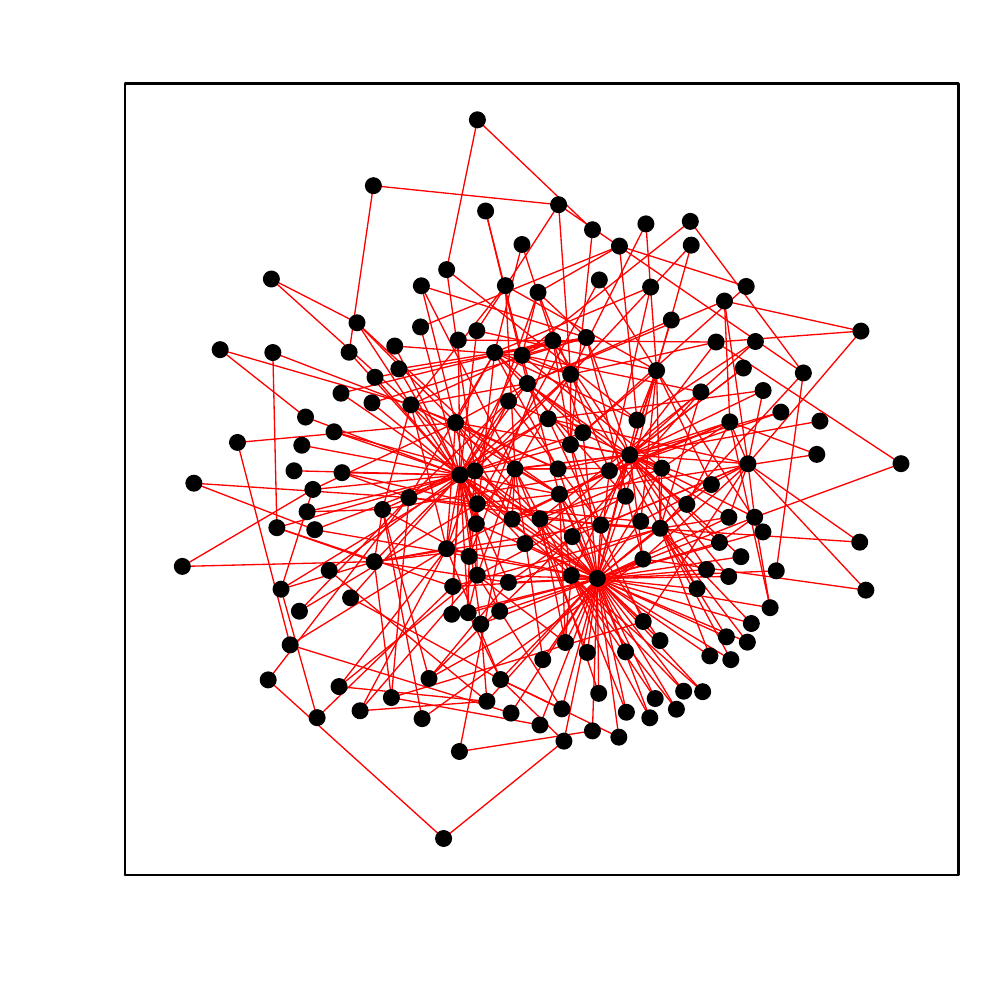}}
	\subfigure[\textsf{netsciecne}.]{\includegraphics[scale=0.45]{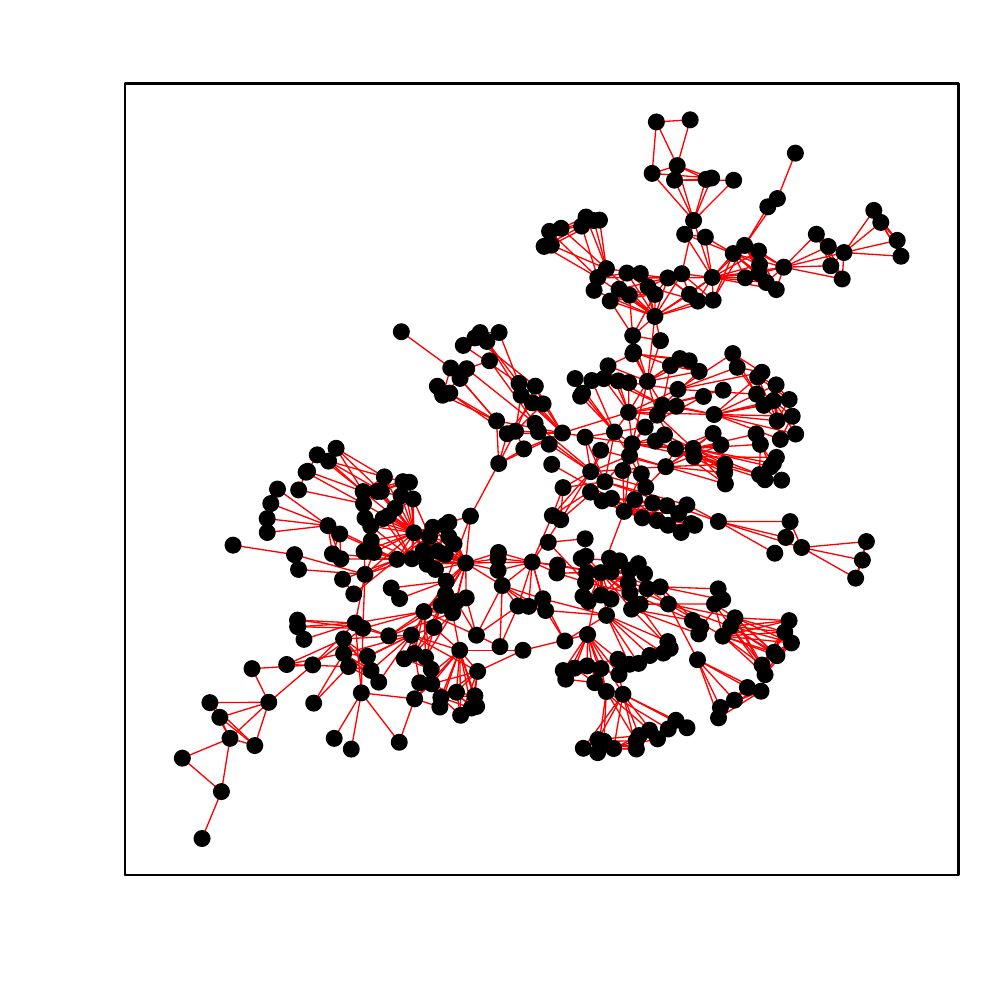}}
	\caption{\footnotesize{Graphs for three selected networks from Table \ref{tab_datasets}.}} 
	\label{fig_networks}
\end{figure}

In order to compare the ability of distance, class, and eigen models to predict missing links, we evaluate their out-of-sample predictive performance through an exhaustive cross-validation experiment under a range of latent dimensions, on 21 networks exhibiting different kinds of actors, sizes, and relations (see Table \ref{tab_datasets} for details about these datasets, which are freely  available on-line. See for example, \url{http://networkrepository.com/}, \url{http://www-personal.umich.edu/~mejn/netdata/}, \url{http://vlado.fmf.uni-lj.si/pub/networks/data/ucinet/ucidata.htm}, and links there in.). Graphs for three selected networks are shown in Figure \ref{fig_networks}.

We fit distance and class models following the same specification given in the previous sections. Now, for the eigen model, we assume that the sampling distribution is given by
$$
y_{i,i'}\mid\zeta,\uv_{i},\uv_{i'}, \LAM \simind \Ber\left( \expit\left(\zeta + \uv_{i}^\trans \LAM \uv_{i'} \right) \right)\,,
$$
where $\uv_i=(u_{i,1},\ldots,u_{i,K})$ is a vector of latent characteristics in $\reals^K$ and $\LAM = \diag{\lambda_1,\ldots,\lambda_K}$ is a diagonal matrix of size $K\times K$, which implies that $ \uv_{i}^\trans \LAM \uv_{i'} = \sum_{k=1}^K\lambda_ku_{i,k}u_{i'k}$ is a quadratic form where $\lambda_k$ weights the contribution of each latent dimension (positively of negatively) to the plausibility of observing an edge between actors $i$ and $i'$. Following the same prior formulation given for distance models, we let $\zeta\mid\ome^2 \sim \Nor(0,\ome^2)$ and $\uv_i\mid\sig^2\simiid \Nor(\zerov,\sig^2\,\I)$, along with $\ome^2 \sim \IGamd(a_\ome,b_\ome)$ and $\sig^2\sim\IGamd(a_\sig,b_\sig)$. We complete the specification by assuming that $\lambda_k\mid\kappa^2 \simiid \Nor(0,\kappa^2)$, where $\kappa^2 \sim  \IGamd(a_{\kappa}, b_{\kappa})$. Lastly, vaguely uninformative priors that have proven to work well in practice are obtain by setting $a_\ome=a_\sig=a_\kappa=2$ and $b_\ome=b_\sig=b_\kappa=3$.

\begin{table}[!htb]
	\centering
	\begin{tabular}{c|ccc|ccc|ccc}  
		\hline
		Network  & \multicolumn{3}{|c|}{\textsf{jazz}} & \multicolumn{3}{|c|}{\textsf{gen}} & \multicolumn{3}{|c}{\textsf{netscience}} \\
		\hline
		$K$ & \textsf{dist} & \textsf{class} & \textsf{eigen} & \textsf{dist} & \textsf{class} & \textsf{eigen} & \textsf{dist} & \textsf{class} & \textsf{eigen}\\ 
		\hline
		2 & 0.914          & 0.721          & 0.910          & 0.596          & 0.779          & 0.723          & \textbf{0.950} & 0.670 & 0.845 \\
		4 & 0.949          & \textbf{0.749} & \textbf{0.940} & 0.668          & \textbf{0.822} & 0.727          & 0.944          & 0.747 & \textbf{0.957} \\
		8 & \textbf{0.971} & 0.742          & 0.876          & \textbf{0.742} & 0.822          & \textbf{0.785} & 0.944          & \textbf{0.820} & 0.849 \\
		\hline
	\end{tabular}
	\caption{{\footnotesize Average AUC values to assess the predictive performance of distance, class, and eigen models, using three selected networks provided in Table \ref{tab_datasets}.}}\label{tab_auc3} 
\end{table}

\begin{figure}[!htb]
	\centering
	\includegraphics[scale=0.9]{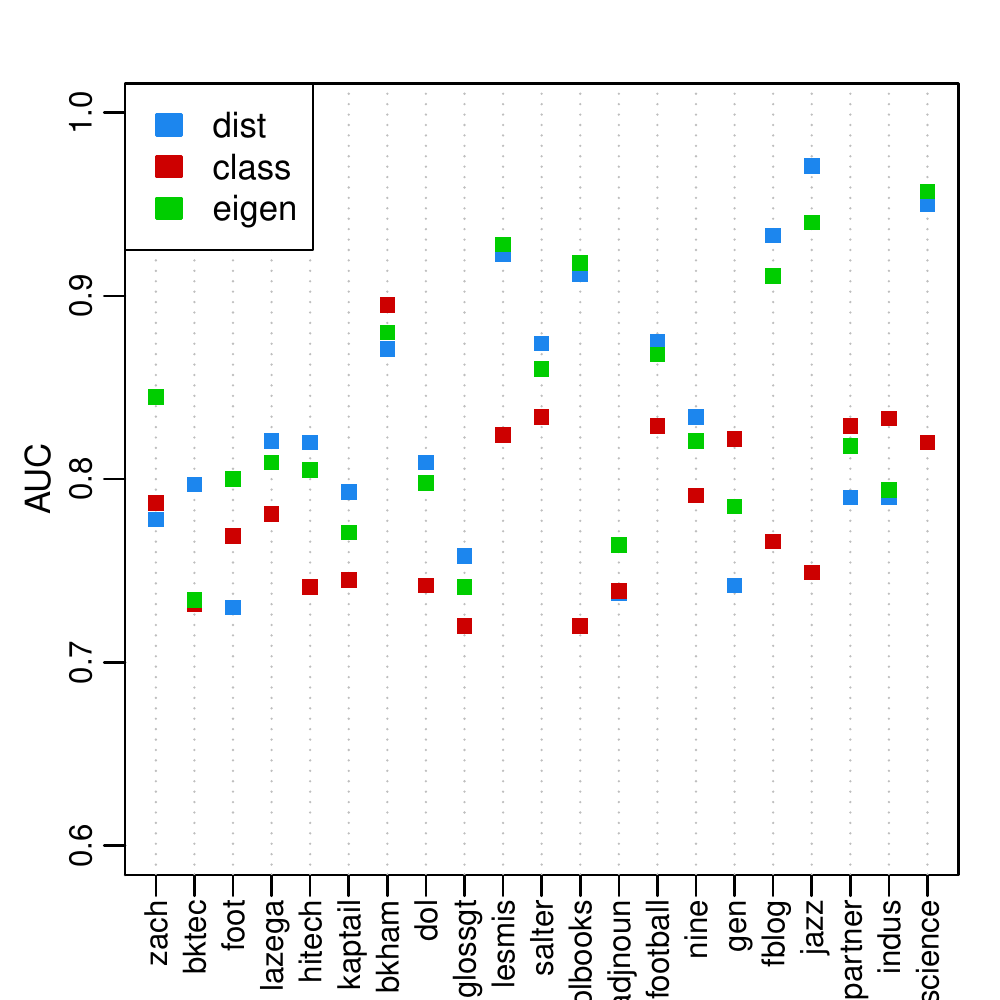}
	\caption{\footnotesize{AUC values for distance, class, and eigen models using each network provided in Table \ref{tab_datasets}. These results correspond to the smallest value of $K$ that maximizes the AUC.}} 
	\label{fig_auc}
\end{figure}

Thus, for each combination of model, dataset, and latent dimension $K\in\{2,4,8\}$, we run a 5-fold cross validation experiment as follows: First, we randomly divide the data into five sets of roughly equal size. Next, for each set $s$, we fit the model conditional on $\{y_{i,i'}:(i,i')\notin s\}$, and for each $y_{k,\ell}$ assigned to $s$, we compute $\expec{y_{k,\ell}\mid\{y_{i,i'}:(i,i')\notin s\}}$, the posterior predictive mean of $y_{k,\ell}$ using all the data not in $s$. Then, using such predictions, we construct a binary classifier to obtain the corresponding receiver operating characteristic (ROC) curve. Lastly, we quantify the predictive performance of each ROC curve through the area under the curve (AUC). In this context, the AUC is a measure of how well a given model is capable of predicting missing links (the higher the AUC, the better the model is at predicting 0s as 0s and 1s as 1s). In every case, inferences are based on $50,000$ samples of the posterior distribution after a burn-in period of 10,000 iterations, by following the corresponding MCMC algorithm provided in Appendix \ref{app_mcmc}.

\begin{figure}[!htb]
	\centering
	\setlength{\tabcolsep}{3pt}
	\begin{tabular}{cccc}
		&  \;\;\; \textsf{jazz}   &  \;\;\; \textsf{gen}   &  \;\;\;\; \textsf{netscience}  \\
		\begin{sideways} \;\;\;\;\;\;\;\;\;\;\;\;\;\;\;\; Density \end{sideways}   &
		\includegraphics[scale=0.4]{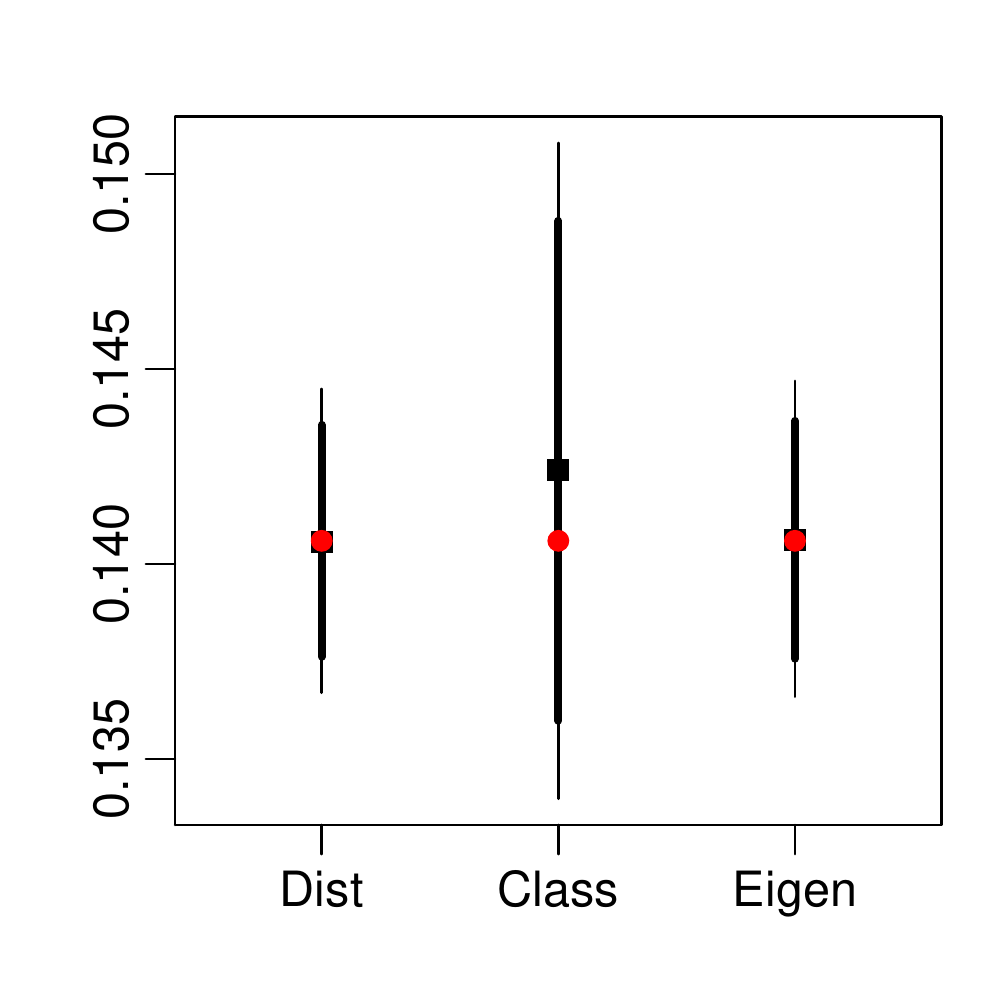}   &
		\includegraphics[scale=0.4]{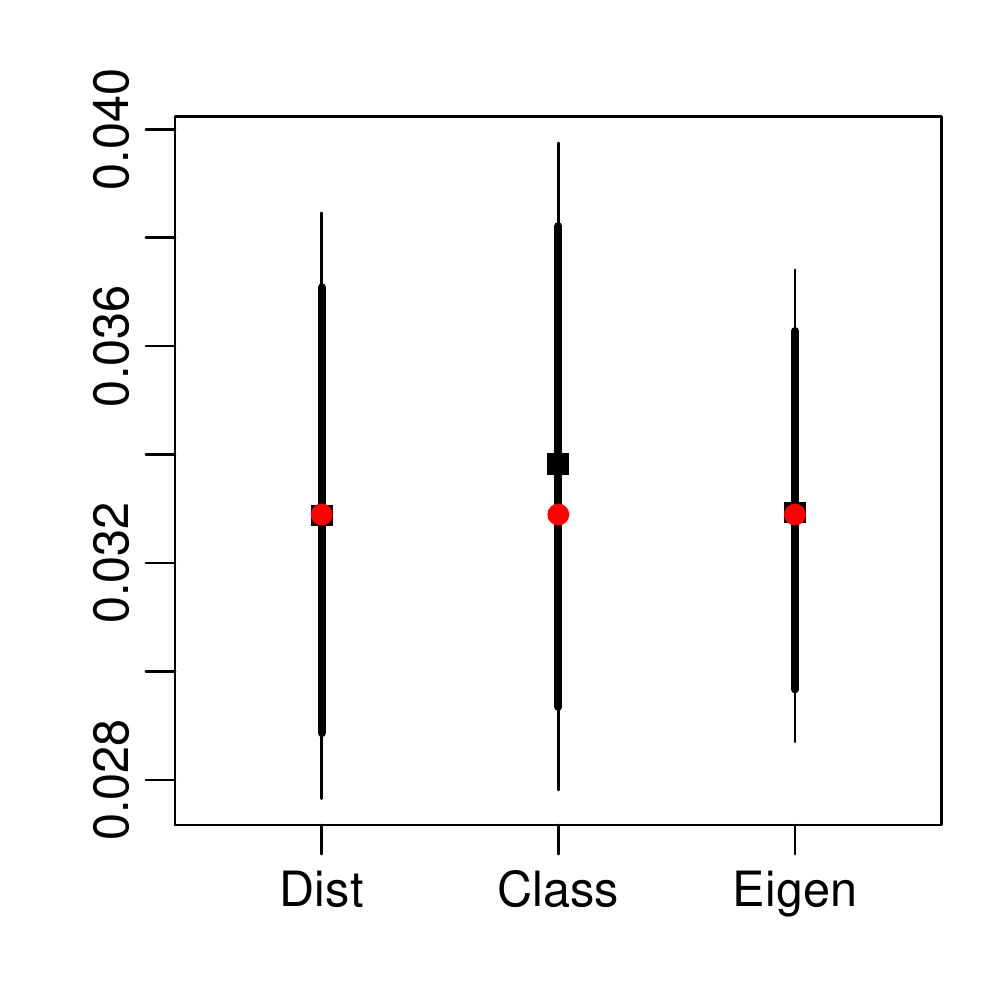}   &
		\includegraphics[scale=0.4]{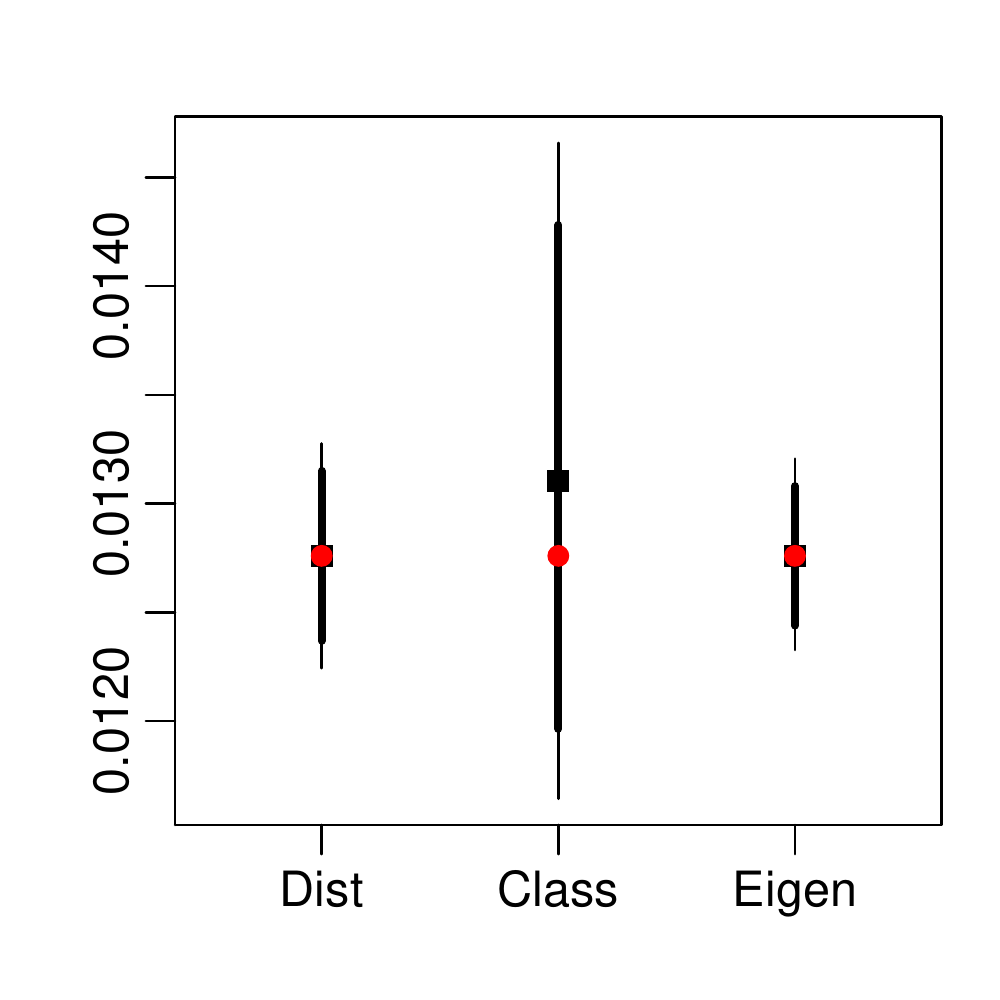}  \\
		\begin{sideways} \;\;\;\;\;\;\;\;\;\;\;\;\; Transitivity \end{sideways}   &
		\includegraphics[scale=0.4]{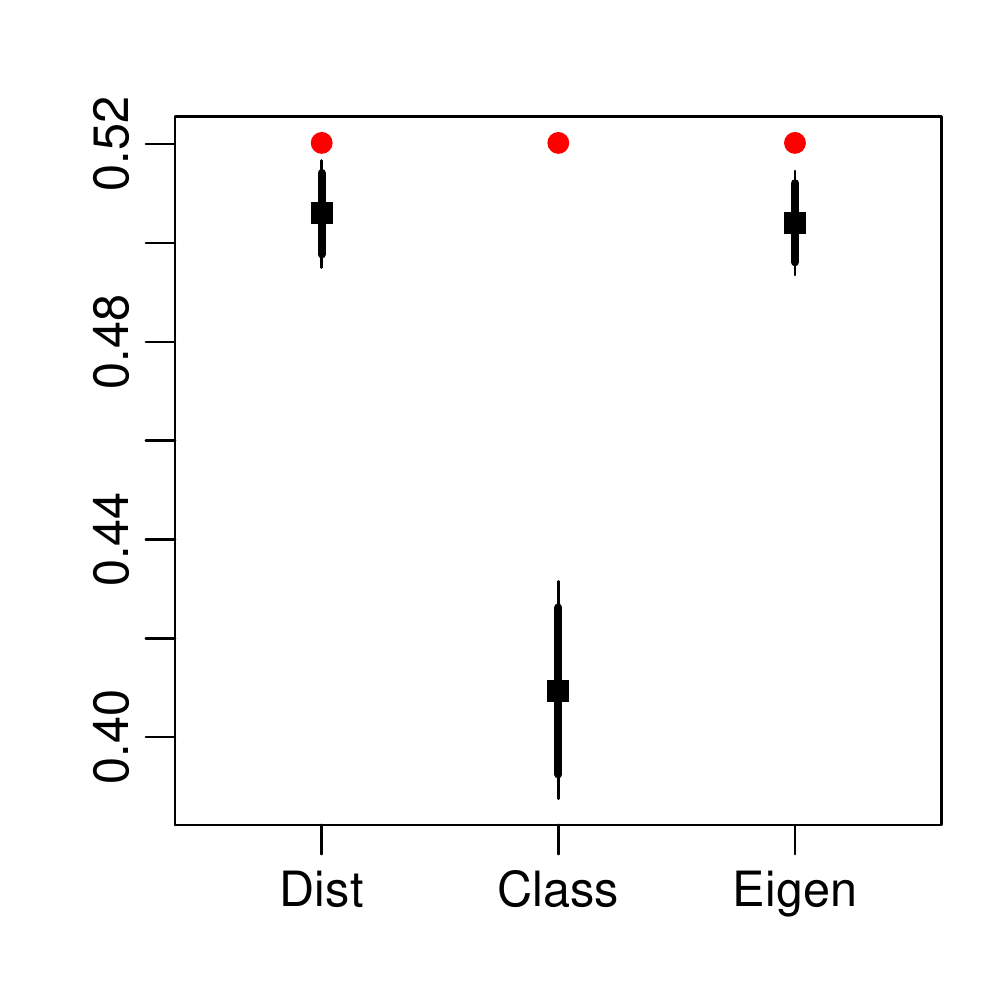}   &
		\includegraphics[scale=0.4]{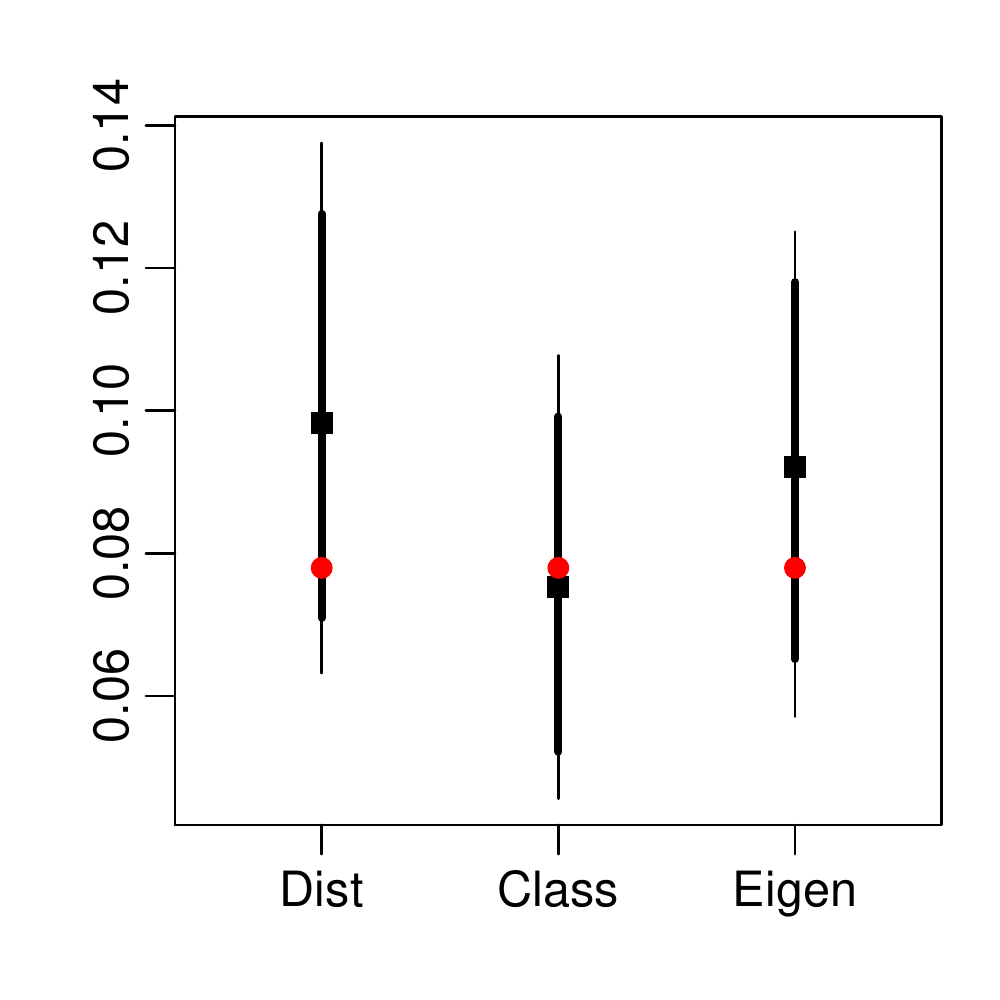}   &
		\includegraphics[scale=0.4]{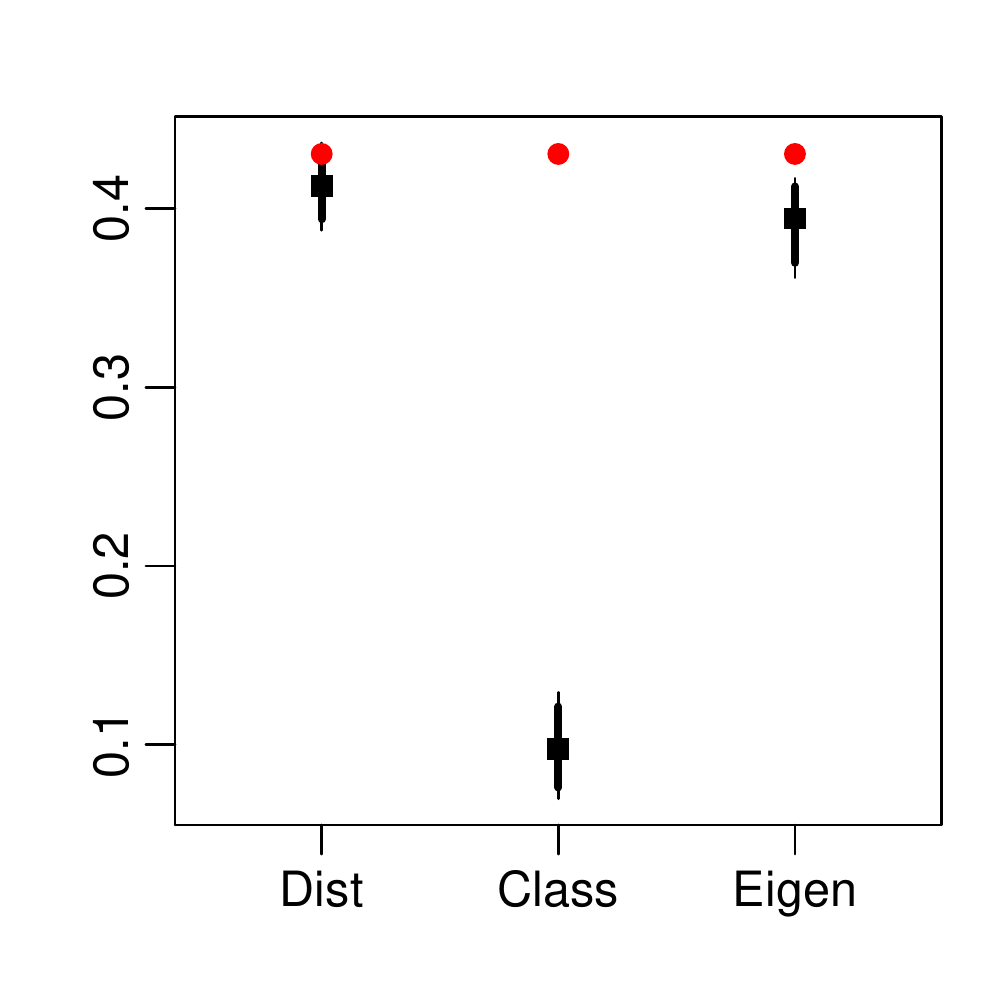}  \\
		\begin{sideways} \;\;\;\;\;\;\;\;\;\;\;\; Assorativity \end{sideways}   &
		\includegraphics[scale=0.4]{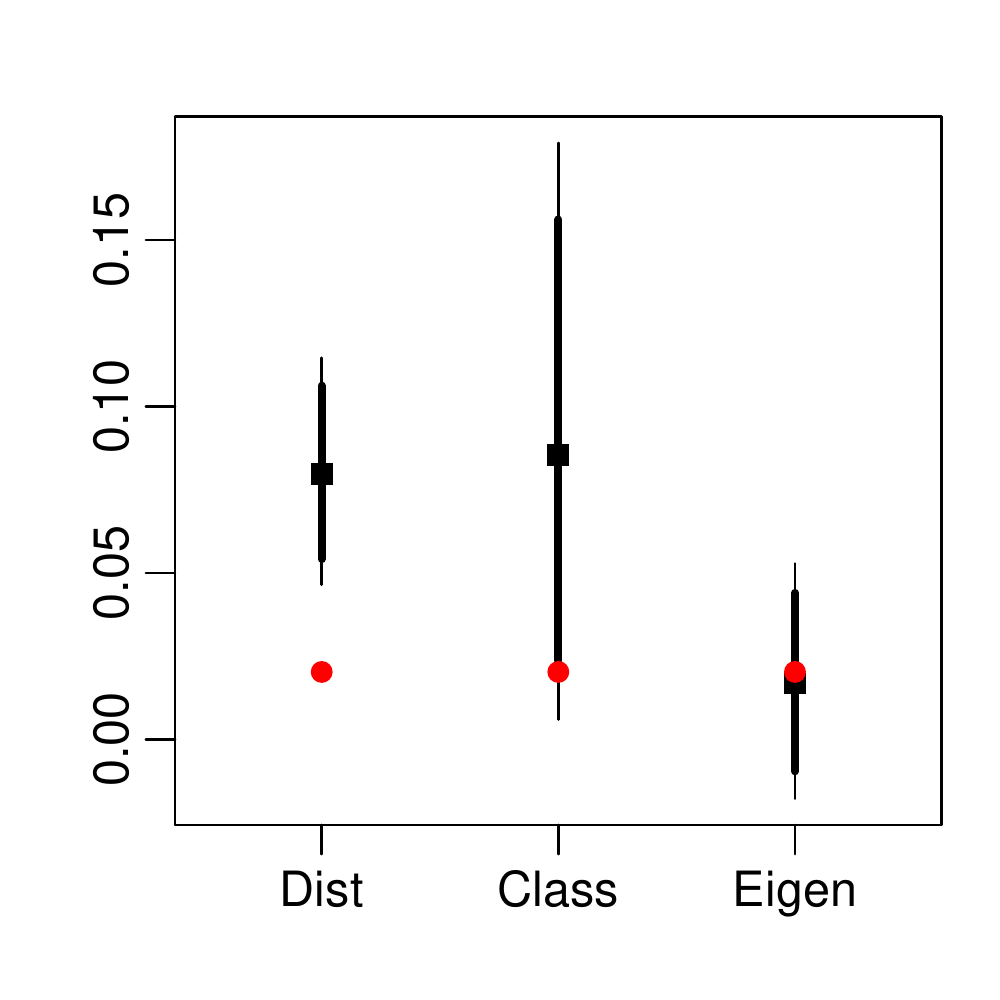}   &
		\includegraphics[scale=0.4]{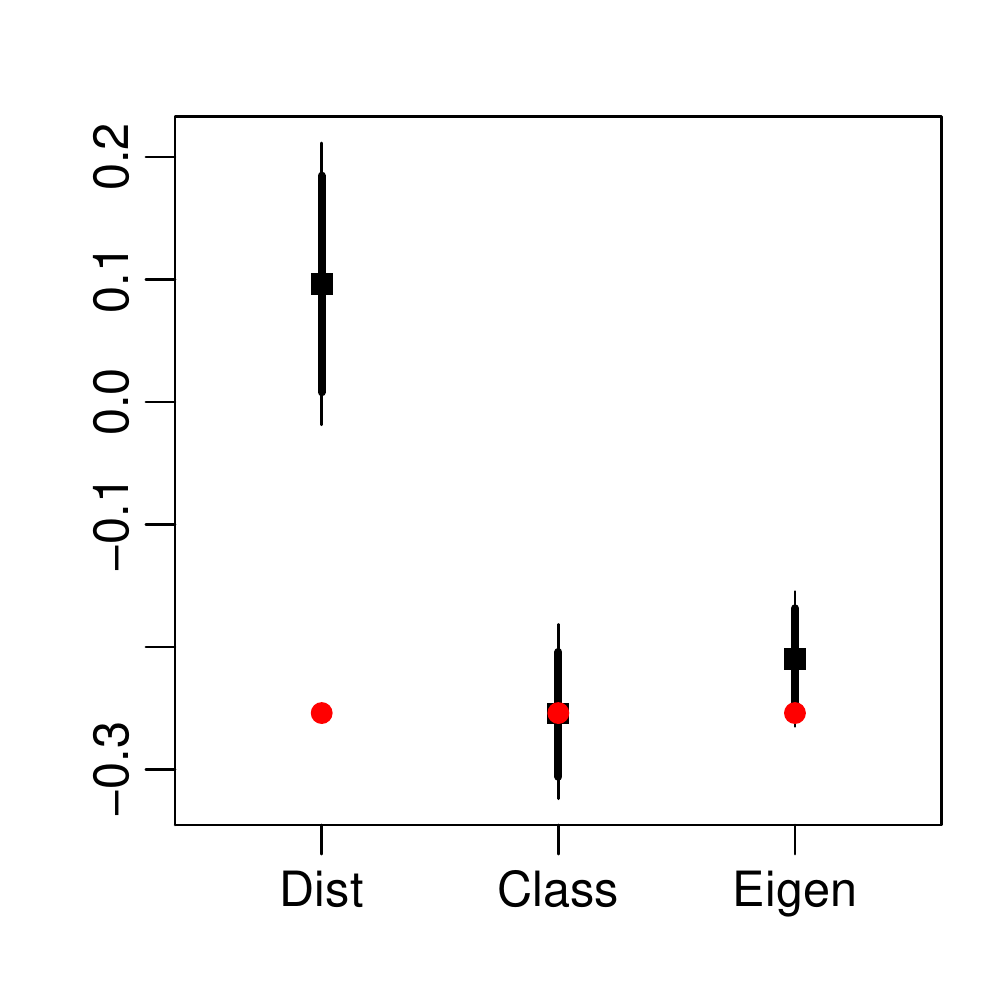}   &
		\includegraphics[scale=0.4]{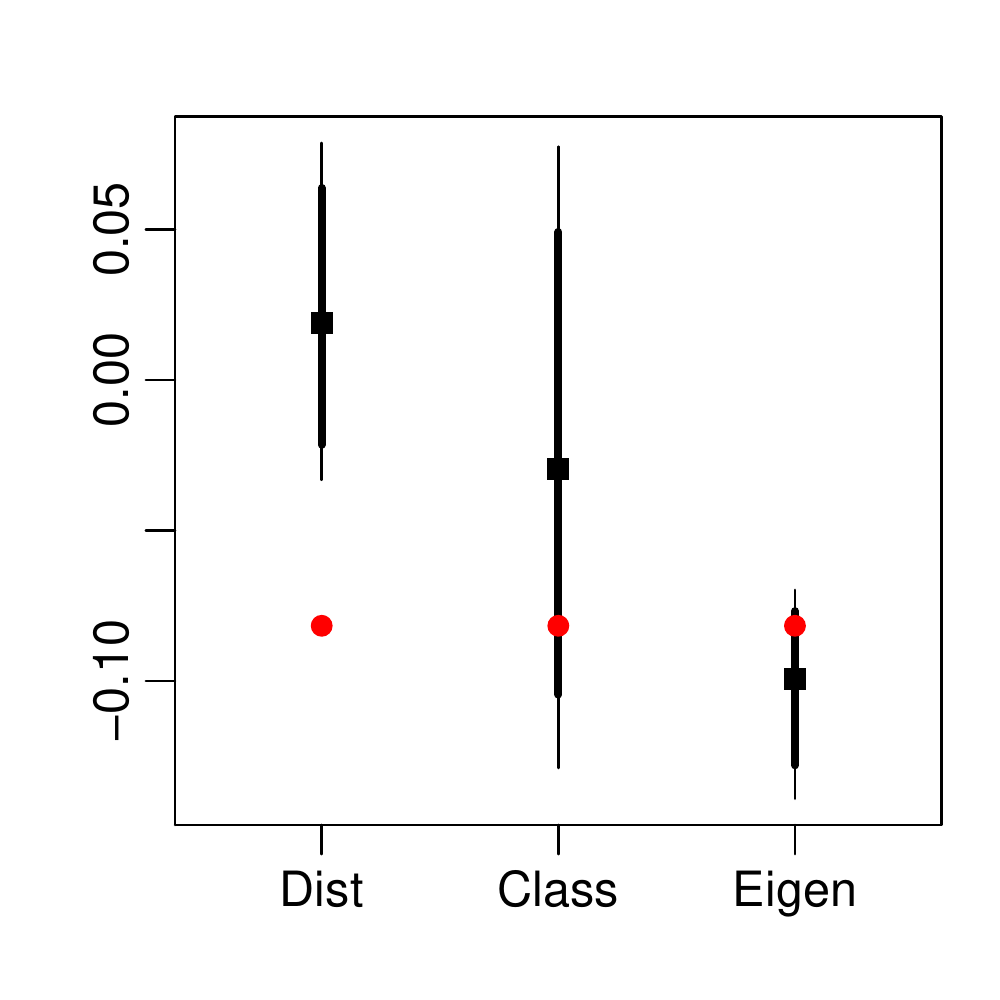}  \\
	\end{tabular}
	\caption{\footnotesize{ Posterior mean (black square) along with 95\% and 99\% credible intervals corresponding to the empirical distribution of test statistics for replicated data along with the observed value (red bullet) in three selected networks. }} 
	\label{fig_stats}
\end{figure}

For three selected networks, we report our findings in Table \ref{tab_auc3}. In addition, Figure \ref{fig_auc} displays the results for the smallest value of $K$ that maximizes the AUC for all the datasets in Table \ref{tab_datasets}. We note that there is no such a thing as a ``best'' model in terms of prediction. A model in particular is more adequate for a given network than another depending on the network's structural features. Distance models have an outstanding predictive performance for those networks with predominant values of transitivity, as well as class models do for those networks exhibiting substantial assortativity levels. 
As expected, eigen models tend to behave quite well predicting missing links under several scenarios since they generalize both distance and class models, but the opposite is not true \citep[Sec. 2.2]{hoff-2008}. Lastly, from Table \ref{tab_datasets}, it is quite evident that the choice of $K$ is key for assuring model performance.

Next, in the same spirit of \citet[Chap. 6]{gelman-2014-information} and \citet[Chap. 4]{kolaczyk2020statistical}, we replicate pseudo-data from all three fitted models and calculate a battery of summary statistics (in our case, density, transitivity, over the selected networks) for each sample from the posterior distribution. This allows us to generate an estimate of the posterior predictive distribution of the summaries, which can then be compared against the value observed in the original sample (Figure \ref{fig_stats}). We see that all models are able to capture the density of each network, although class models are more uncertain in regard with the corresponding estimate. Furthermore, distance and eigen models are clearly capable of reproducing transitivity patterns, unlike class models that underestimate such feature. On the contrary, distance models tend to overestimate assortativity values, whereas both class and eigen models successfully register this characteristic. Not surprisingly, eigen models are able to capture most of the structural features of the data and have less uncertainty attached to their estimates.

\begin{table}[H]
	\centering
	\begin{tabular}{lccc}
		\hline
		net & \textsf{Dist} & \textsf{Class} & \textsf{Eigen} \\ 
		\hline
		\textsf{zach}     & 378.7   & 377.5    & \textbf{296.8}   \\ 
		\textsf{bktec}    & \textbf{565.1}   & 636.0    & 592.9   \\ 
		\textsf{foot}     & 509.9   & 552.0    & \textbf{434.3}   \\ 
		\textsf{lazega}   & 454.9   & 545.5    & \textbf{452.5}   \\ 
		\textsf{hitech}   & \textbf{387.7}   & 480.5    & 390.0   \\ 
		\textsf{kaptail}  & \textbf{603.6}   & 721.7    & 604.7   \\ 
		\textsf{bkham}    & 591.2   & 579.9    & \textbf{454.3}   \\ 
		\textsf{dol}      & \textbf{739.2}   & 958.1    & 834.8   \\ 
		\textsf{glossgt}  & \textbf{775.1}   & 893.7    & 805.0   \\ 
		\textsf{lesmis}   & 999.8   & 1,667.2  & \textbf{919.2}   \\ 
		\textsf{salter}   &\textbf{2,200.3} & 2,789.7  & 2,275.7 \\ 
		\textsf{polbooks} & \textbf{2,003.2} & 2,904.4  & 2,011.2 \\ 
		\textsf{adjnoun}  & 2,855.0 & 2,845.8  & \textbf{2,601.5} \\ 
		\textsf{football} & \textbf{2,700.9} & 3,759.7  & 3,351.4 \\ 
		\textsf{nine}     & 1,101.5 & 1,384.0  & \textbf{1,070.2} \\ 
		\textsf{gen}      & 3,455.6 & 3,134.5  & \textbf{2,813.5} \\ 
		\textsf{fblog}    & \textbf{6,431.7} & 9,283.3  & 7,014.4 \\ 
		\textsf{jazz}     & 8,938.6 & 13,590.5 & \textbf{8,707.7} \\ 
		\textsf{partner}  & 5,163.2 & 5,488.1  & \textbf{4,624.0} \\ 
		\textsf{indus}    & 5,159.7 & 5,096.4  & \textbf{4,725.8} \\ 
		\textsf{science}  & \textbf{4,108.5} & 9,419.4  & 7,364.3 \\ 
		\hline
	\end{tabular}
	\caption{\footnotesize{ Values of WAIC for distance, class, and eigen models using each network provided in Table \ref{tab_datasets}. These results correspond to the smallest value of $K$ that minimizes the WAIC.}} 
	\label{tab_waic}
\end{table}

Finally, in order to asses the goodness-of-fit of each model, we complement the results presented above by considering measures that account for both model fit and model complexity. Such measures also serve as a tool for model selection, since the value of latent dimension $K$ can potentially play a critical role in the results. The network literature has largely focused on the Bayesian Information Criteria (BIC) as a tool for model selection, e.g. \citet{hoff-2005}, \citet{handcock-2007} and \citet{airoldi-2009}).  However, BIC is typically inappropriate for hierarchical models since the hierarchical structure implies that the effective number of parameters will typically be lower than the actual number of parameters in the likelihood.  Two alternatives to BIC that address this issue are the Deviance Information Criterion \citep[DIC]{spiegelhalter2002bayesian,spiegelhalter-2014},
$$\DIC(K)  = - 2\log p ( \Y \mid \hat{\UPS}_K ) + 2p_{\DIC}\,,$$
with
$p_{\DIC} =   2\log p ( \Y \mid \hat{\UPS}_K ) - 2\expec{\log p \left(\Y \mid \UPS_{K} \right) }$,
and the Watanabe-Akaike Information Criterion \citep[WAIC]{watanabe2010asymptotic,watanabe2013widely},
$$
\WAIC(K)  = -2\,\sum_{i<i'}\log \expec{ p \left( y_{i,i'} \mid {\UPS}_K \right) }  + 2\,p_{\WAIC}\,,
$$
with $p_{\WAIC} = 2\sum_{i<i'} \big\{ \log\expec{p \left( y_{i,i'}|\UPS_{K} \right)} - \expec{\log p \left( y_{i,i'}|\UPS_K \right)} \big\}$,
where $\hat{\UPS}_K$ denotes the posterior mean of model parameters assuming that the dimension of the social space is $K$, and $p_{\DIC}$ and $p_{\WAIC}$ are penalty terms accounting for model complexity. Note that in the previous expressions all expectations, which are computed with respect to the posterior distribution, can be approximated by averaging over Markov chain Monte Carlo (MCMC) samples (see Section \ref{sec_computation} for details). A key advantage of the WAIC criteria is its invariance to reparameterizations, which makes it particularly helpful for models (such as ours) with hierarchical structures, for which the number of parameters increases with sample size \citep{gelman-2014-information,spiegelhalter-2014}. Table \ref{tab_auc3} presents the results for the smallest value of $K$ that minimizes the AUC for all the datasets in Table \ref{tab_datasets}. We see that distance models and eigen models provide the best fit according to the WAIC.

Here, we have adopted a standard procedure to selecting the dimension of the latent by means of the WAIC. However, the latent dimension can also be treated directly as a model parameter by placing a prior distribution on it, in the same spirit of \citet{green2009reversible}. On the other hand, a similar approach discussed in \citep{guhaniyogi2018joint}, which can be understood as a truncation of a non-parametric process, could be incorporated here for selecting the latent dimension. Nonetheless, based on the evidence provided by \citet{guhaniyogi2018joint}, the results are likely to be quite similar.

\section{Discussion}\label{sec_discussion}

Our fundings show that the performance of the latent space models is case-specific, in terms of both goodness-of-fit and prediction. Each model has weaknesses and strengths. For example, class models are very suitable for networks exhibiting high levels of clustering, whereas distance models are preferred to represent major degrees of transitivity. However, eigen models seem to behave very well under a great variety of scenarios, which is quite logical since it generalizes (qualitatively) class and distance models \cite{hoff-2008}.

Latent space models have proven to be extremely in all sorts of applications involving social network data due to their flexibility and interpretability. Some applications and extensions include modeling of multilayer networks \citep{salter2017latent,durante2018bayesian}, cognitive social structures \citep{sosa2017latent}, dynamic networks \citep{han2015consistent,hoff2015multilinear,sewell2015latent}, record linkage \citep{sosa2018record,sosa2019bayesian}, and community detection \citep{regueiro2017scalable,paez2019hierarchical}, among many others, with all sort of implications and ramifications, e.g., fast computation for ``big networks'' \citep{raftery2012fast,salter2013variational}). For more reviews in special topics related to latent space models, we refer the reader to \citet{sweet2013hierarchical}, \citet{rastelli2015properties}, \citet{kim2018review}, and \citet{minhas2019inferential}.

As a final note, we acknowledge that there are available many extensions of the basic latent models presented here, which are quite common in the network literature. Such modifications and extensions include incorporation of covariates (with its many variants) and popularity parameters, for instance. See \cite{raftery2017comment} for some ideas in this regard.

    \bibliography{references}

    \appendix

\appendix

\section{MCMC algorithms}\label{app_mcmc}

Our MCMC algorithm iterates over the model parameters $\UPS$. Where possible we sample from the full conditional posterior distributions as in Gibbs sampling; otherwise we use adaptive versions of either Metropolis-Hastings or Hamiltonian Monte Carlo steps. Alternatively, in the same spirit of \cite{albert-1993}, Polya-Gamma random variables can be introduced in order to facilitate computation \citep{polson2013bayesian}.

\subsection{Distance model}\label{app_mcmc_distance}

The joint posterior distribution is given by:
\begin{align*}
p(\UPS\mid\Y) &= p(\Y\mid \zeta, \{\uv_i\})\,p(\{\uv_i\}\mid \sigma^2)\,p(\sigma^2)\,p(\zeta\mid\omega^2)\,p(\omega^2)\\
&\propto\prod_{i=1}^{I-1}\prod_{i'=i+1}^I \theta_{i,i'}^{y_{i,i'}}(1-\theta_{i,i'})^{1-y_{i,i'}} \times
\prod_{i=1}^I (\sig^2)^{-K/2}\,\ex{-\tfrac{1}{2\sig^2}\,\|\uv_i\|^2} \\
&\hspace{0.5cm} \times(\sig^2)^{-(a_\sig+1)}\,\ex{-\frac{b_\sig}{\sig^2}}
\times (\ome^2)^{-1/2} \ex{-\tfrac{1}{2\ome^2}\,\zeta^2}  \times (\ome^2)^{-(a_\ome+1)}\,\ex{-\frac{b_\ome}{\ome^2}}\,,
\end{align*}
where $\theta_{i,i'}=\expit(\zeta- \|\uv_i - \uv_{i'}\|)$ and
$\UPS= (\uv_1,\ldots,\uv_I, \zeta, \sigma^2, \ome^2)$ is the set of model parameters.

For a given set of hyperparameters $(a_{\sig}, b_{\sig}, a_{\ome}, b_{\ome})$, the algorithm proceeds by generating a new state $\UPS^{(b+1)}$ from a current state $\UPS^{(b)}$, $b=1,\ldots,B$, as follows:
\begin{enumerate}
	\item Sample $\uv_i^{(b+1)}$, $i = 1,\ldots,I$, according to a Metropolis--Hastings Algorithm, considering the full conditional distribution:
	\begin{align*}
	p(\uv_i\mid\rest)
	&\propto\prod_{i'=i+1}^I \theta_{i,i'}^{y_{i,i'}}(1-\theta_{i,i'})^{1-y_{i,i'}}\times
	\prod_{i'=1}^{i-1} \theta_{i',i}^{y_{i',i}}(1-\theta_{i',i})^{1-y_{i',i}} \times\ex{-\tfrac{1}{2\sig^2}\,\|\uv_i\|^2}\,.
	\end{align*}
	\item Sample $\zeta^{(b+1)}$ according to a Metropolis--Hastings Algorithm, considering the full conditional distribution:
	\begin{align*}
	p(\zeta\mid\rest)
	&\propto
	\prod_{i=1}^{I-1}\prod_{i'=i+1}^I \theta_{i,i'}^{y_{i,i'}}(1-\theta_{i,i'})^{1-y_{i,i'}}\times \ex{-\tfrac{1}{2\ome^2}\,\zeta^2}.
	\end{align*}
	
	\item Sample $(\si^2)^{(b+1)}$ from
	$p(\sig^2\mid\rest) = \IGamd\le(\sig^2 \mid a_\sig + \tfrac{I\,K}2, b_\sig + \tfrac12 \textstyle\sum_{i=1}^I\|\uv_i\|^2 \ri)\,.$
	
	\item Sample $(\ome^2)^{(b+1)}$ from
	$p(\ome^2\mid\rest) = \IGamd\le(\ome^2 \mid a_\omega + \tfrac12, b_\omega + \tfrac12 \zeta^2 \ri)\,.$
\end{enumerate}

\subsection{Class model}\label{app_mcmc_class}

The joint posterior distribution is given by:
\begin{align*}
p(\UPS\mid\Y) &= p(\Y\mid\{\xi_i\},\{\eta_{k,\ell}\})\,p(\{\eta_{k,\ell}\}\mid\zeta,\tau^2)\,p(\zeta)\,p(\tau^2)\,p(\{\xi_i \}\mid\omev)\,p(\omev\mid\alpha)\,p(\alpha)\\
&\propto\prod_{i=1}^{I-1}\prod_{i'=i+1}^I \theta_{i,i'}^{y_{i,i'}}(1-\theta_{i,i'})^{1-y_{i,i'}}
\times \ex{-\tfrac{1}{2\sig_\zeta^2}(\zeta-\mu_\zeta)^2} \times (\tau^2)^{-(a_\tau-1)}\,\ex{-\frac{b_\tau}{\tau^2}} \\
&\hspace{0.5cm} \times \prod_{k=1}^K\prod_{\ell=k}^K (\tau^2)^{-1/2}\,\ex{-\tfrac{1}{2\tau^2}(\eta_{k,\ell} - \zeta)^2} \times \prod_{i=1}^I\prod_{k=1}^K\omega_k^{[\xi_i=k]} \times \frac{\Gamma\left(\tfrac{\al}{K}\right)^K}{\Gamma(\alpha)}\,\prod_{k=1}^K \ome_k^{\tfrac{\al}{K}-1}\\
&\hspace{0.5cm}\times \alpha^{a_\al-1}\,\ex{-b_\al\,\al}\,,
\end{align*}
where $\theta_{i,i'}=\expit(\eta_{\phi(\xi_i,\xi_{i'})})$ and 
$\UPS=(\eta_{1,1},\eta_{1,2},\ldots,\eta_{K,K}, \xi_1,\ldots,\xi_I,\omega_1,\ldots,\omega_K,\zeta,\tau^2,\alpha)$
is the set of model parameters.

For a given set of hyperparameters $(\mu_\zeta,\sigma^2_\zeta,a_\tau,b_\tau,a_\al,b_\al)$, the algorithm proceeds by generating a new state $\UPS^{(b+1)}$ from a current state $\UPS^{(b)}$, $b=1,\ldots,B$, as follows:

\begin{enumerate}
	
	\item Sample $\eta_{k,\ell}^{(b+1)}$, $\ell = k,\ldots,K$ and $k = 1,\ldots,K$, according to a Metropolis--Hastings Algorithm, considering the full conditional distribution:
	\begin{align*}
	\log p(\eta_{k,\ell}\mid\rest) 
	&\propto 
	s_{k,\ell}\log(\expit\eta_{k,\ell}) + (n_{k,\ell}-s_{k,\ell})\log(1-\expit\eta_{k,\ell}) - \frac{1}{2\tau^2}(\eta_{k,\ell} - \zeta)^2 \\
	&= s_{k,\ell}\,\eta_{k,\ell} - n_{k,\ell} \log(1 + \exp\,\eta_{k,\ell}) - \frac{1}{2\tau^2}(\eta_{k,\ell} - \zeta)^2\,,
	\end{align*}
	where $s_{k,\ell} = \sum_{\mathcal{S}_{k,\ell}} y_{i,i'}$ and $n_{k,\ell} = \sum_{\mathcal{S}_{k,\ell}} 1$, with $\mathcal{S}_{k,\ell} = \{(i,i'):i<i'\text{ and }\phi(\xi_i,\xi_{i'})=(k,\ell)\}$. 
	
	\item Sample $\xi_i^{(b+1)}$, $i=1,\ldots,I$, from a categorical distribution on $\{1,\ldots,K\}$, such that:
	$$
	\pr{\xi_i=k\mid\rest}\propto \omega_k
	\times\prod_{i'=i+1}^I \eta_{\phi(k,\xi_{i'})}^{y_{i,i'}}(1-\eta_{\phi(k,\xi_{i'})})^{1-y_{i,i'}}
	\times\prod_{i'=1}^{i-1} \eta_{\phi(\xi_{i'},k)}^{y_{i',i}}(1-\eta_{\phi(\xi_{i'},k)})^{1-y_{i',i}}\,.
	$$
	
	\item Sample $\omev^{(b+1)}$ from $p(\omev\mid\rest) = \Dir\left(\omev\mid\tfrac{\al}{K}+n_1,\ldots,\tfrac{\al}{K}+n_K\right)$, where $n_k$ is the number of actors in cluster $k\in\{1,\ldots,K\}$.
	
	\item Sample $\zeta^{(b+1)}$ from $\Nor(m,v^2)$, where
	$$v^2 = \left(\frac{1}{\sigma^2_\zeta} + \frac{K(K+1)/2}{\tau^2}\right)^{-1} \qquad\text{and}\qquad m =	v^2\,\left( \frac{\mu_\zeta}{\sigma^2_\zeta} + \frac{1}{\tau^2}\sum_{k=1}^K\sum_{\ell=k}^K\eta_{k,\ell} \right)\,.$$
	
	\item Sample $(\si^2)^{(b+1)}$ from
	$$p(\sig^2\mid\rest) = \IGamd\le(\sig^2 \mid a_\tau + \tfrac{K(K+1)}4, b_\tau + \tfrac12\sum_{k=1}^K\sum_{\ell=k}^K (\eta_{k,\ell}-\zeta)^2  \ri)\,.$$
	
	\item Sample $\alpha^{(b+1)}$ according to a Metropolis--Hastings Algorithm, considering the full conditional distribution:
	\begin{align*}
	\log p(\alpha\mid\rest)
	&\propto
	\log\Gamma(\alpha) - K\log\Gamma(\alpha/K) + \frac{\al}{K}\sum_{k=1}^K\log\omega_k - (a_\beta-1)\log\al - b_\al\,\al\,.
	\end{align*}
	
\end{enumerate}

\subsection{Eigen model}\label{app_mcmc_eigen}

The joint posterior distribution is given by:
\begin{align*}
p(\UPS\mid\Y) &= p(\Y\mid\zeta,\{\uv_i \}, \{\lambda_k\} )\,p(\{\uv_i \}\mid\sigma^2)\,p(\sigma^2)\,p(\{\lambda_k \}\mid\kappa^2)\,p(\kappa^2)\,p(\zeta\mid\omega^2)\,p(\omega^2)\\
&\propto\prod_{i=1}^{I-1}\prod_{i'=i+1}^I \theta_{i,i'}^{y_{i,i'}}(1-\theta_{i,i'})^{1-y_{i,i'}} \times
\prod_{i=1}^I (\sig^2)^{-K/2}\,\ex{-\tfrac{1}{2\sig^2}\,\|\uv_i\|^2} \times(\sig^2)^{-(a_\sig+1)}\,\ex{-\frac{b_\sig}{\sig^2}} \\
&\hspace{0.5cm} \times \prod_{k=1}^K (\kap^2)^{-1/2} \ex{-\tfrac{1}{2\kap^2}\,\lambda_k^2} \times(\kap^2)^{-(a_\kap+1)}\,\ex{-\frac{b_\kap}{\kap^2}} \times (\ome^2)^{-1/2} \ex{-\tfrac{1}{2\ome^2}\,\zeta^2}  \\
&\hspace{0.5cm} \times (\ome^2)^{-(a_\ome+1)}\,\ex{-\frac{b_\ome}{\ome^2}}\,,
\end{align*}
where $\theta_{i,i'}=\expit(\zeta + \uv_i \LAM \uv_{i'})$ and 
$\UPS= (\uv_1,\ldots,\uv_I, \lambda_1,\ldots,\lambda_K, \zeta, \sigma^2, \kappa^2, \ome^2)$
is the set of model parameters.

For a given set of hyperparameters $(a_{\sig}, b_{\sig}, a_\kap, b_\kap, a_{\ome}, b_{\ome})$, the algorithm proceeds by generating a new state $\UPS^{(b+1)}$ from a current state $\UPS^{(b)}$, $b=1,\ldots,B$, as follows:

\begin{enumerate}
	\item Sample $\uv_i^{(b+1)}$, $i=1,\ldots,I$, according to a Metropolis--Hastings Algorithm, considering the full conditional distribution:
	\begin{align*}
	p(\uv_i\mid\rest)
	&\propto\prod_{i'=i+1}^I \theta_{i,i'}^{y_{i,i'}}(1-\theta_{i,i'})^{1-y_{i,i'}}\times
	\prod_{i'=1}^{i-1} \theta_{i',i}^{y_{i',i}}(1-\theta_{i',i})^{1-y_{i',i}} \times\ex{-\tfrac{1}{2\sig^2}\,\|\uv_i\|^2}\,.
	\end{align*}
	
	\item Sample $\lambda_k^{(b+1)}$ according to a Metropolis--Hastings Algorithm, considering the full conditional distribution:
	$$
	p(\lambda_k\mid\rest) \prop \prod_{i=1}^{I-1}\prod_{i'=i+1}^I\theta_{i,i'}^{y_{i,i'}}(1-\theta_{i,i'})^{1-y_{i,i'}}\times \ex{-\tfrac{1}{2\kap^2}\,\lambda_k^2}\,.
	$$
	\item Sample $\zeta^{(b+1)}$ according to a Metropolis--Hastings Algorithm, considering the full conditional distribution:
	\begin{align*}
	p(\zeta\mid\rest)
	&\propto
	\prod_{i=1}^{I-1}\prod_{i'=i+1}^I \theta_{i,i'}^{y_{i,i'}}(1-\theta_{i,i'})^{1-y_{i,i'}}\times \ex{-\tfrac{1}{\ome^2}\,\zeta^2}.
	\end{align*}
	
	\item Sample $(\si^2)^{(b+1)}$ from
	$p(\sig^2\mid\rest) = \IGamd\le(\sig^2 \mid a_\sig + \tfrac{I\,K}2, b_\sig + \tfrac12 \textstyle\sum_{i=1}^I\|\uv_i\|^2 \ri)\,.$
	
	\item Sample $(\kap^2)^{(b+1)}$ from
	$p(\kap^2\mid\rest) = \IGamd\le(\kap^2 \mid a_\kap + \tfrac{K}2, b_\kap + \tfrac12 \textstyle\sum_{k=1}^K\lambda_k^2 \ri)\,.$
	
	\item Sample $(\ome^2)^{(b+1)}$ from
	$p(\ome^2\mid\rest) = \IGamd\le(\ome^2 \mid a_\sig + \tfrac12, b_\sig + \tfrac12 \zeta^2 \ri)\,.$
\end{enumerate}

\end{document}